\documentclass[graybox]{svmult}
\usepackage{longtable}
\usepackage{type1cm}        
\usepackage{makeidx}         
\usepackage{graphicx}        
\usepackage{multicol}        
\usepackage[bottom]{footmisc}
\usepackage{newtxtext}       %
\usepackage{newtxmath}       
\makeindex             

\def\lsim{\mathrel{\rlap{\lower4pt\hbox{\hskip1pt$\sim$}}
    \raise1pt\hbox{$<$}}}         
\def\gsim{\mathrel{\rlap{\lower4pt\hbox{\hskip1pt$\sim$}}
    \raise1pt\hbox{$>$}}}         
    
\begin{document}

\title*{Nuclear-powered X-ray millisecond pulsars}
\author{Sudip Bhattacharyya}
\institute{Sudip Bhattacharyya \at Tata Institute of Fundamental Research, Mumbai 400005, India, \email{sudip@tifr.res.in}}
%
%
\maketitle

\label{ch:sudip}
\abstract{Nuclear-powered X-ray millisecond pulsars are the third type of millisecond pulsars, which are powered by thermonuclear fusion processes. The corresponding brightness oscillations, known as {\it burst oscillations}, are observed during some thermonuclear X-ray bursts, when the burning and cooling accreted matter gives rise to an azimuthally asymmetric brightness pattern on the surface of the spinning neutron star. Apart from providing neutron star spin rates, this X-ray timing feature can be a useful tool to probe the fundamental physics of neutron star interior and surface. This chapter presents an overview of the relatively new field of nuclear-powered X-ray millisecond pulsars.}

\section{Introduction}\label{Introduction}

Most of the known millisecond pulsars (MSPs), i.e., rapidly spinning neutron stars, are powered by the stellar spin or rotational kinetic energy 
(see Chap.~1;
\cite{Bhattacharyaetal1991}). There is also a small set of known MSPs, which are powered by the gravitational potential energy released from the matter accreted by the neutron star from its binary companion (see 
Chap.~4;
\cite{PatrunoWatts2012}). In this chapter, however, we will discuss a third set of MSPs, which  are powered by nuclear fusion reactions in the accreted matter accumulated on the neutron star surface \cite{Watts2012}.

Both the second and third types of MSPs are subsets of low-mass X-ray binary (LMXB) systems. In such a system, a neutron star accretes matter from a low-mass object filling its Roche lobe (see 
Chap.~4;
\cite{Bhattacharyaetal1991}). This object could be a main sequence star, an evolved star, a white dwarf or even a brown dwarf (see 
Chap.~7).
An LMXB primarily emits X-rays, as the blackbody temperatures of the inner part of the accretion disk and the neutron star surface are $\sim 1$~keV. Consequently, both types of MSPs, i.e., accretion-powered millisecond X-ray pulsars (AMXPs) and nuclear-powered millisecond X-ray pulsars (NMXPs) are typically observed in X-rays. There is also a partial overlap between these two types of MSPs, i.e., some AMXPs are also NMXPs, and vice versa \cite{Watts2012}.

The three types of pulsations are useful to probe some distinctly different aspects of physics. Therefore, even though NMXPs are much less explored than spin-powered MSPs, the study of the former is a uniquely important field of research by its own right. This is because, apart from providing the neutron star spin rates, which all pulsars do, nuclear-powered pulsations can be useful to probe the supra-nuclear density degenerate core matter of neutron stars (see 
Chap.~9;
\cite{Bhattacharyya2010}) and the stellar surface physics in extreme conditions, such as strong gravity, high magnetic field, intense radiation and rapid stellar spin.

Nuclear-powered pulsations are also called burst oscillations, as this feature is the brightness oscillations observed during thermonuclear X-ray bursts originated on the neutron star surface. It is, therefore,  imperative to briefly describe these bursts, in order to understand NMXPs.

\subsection{Thermonuclear X-ray bursts}\label{bursts}

\begin{figure}[t]
\sidecaption[t]
\includegraphics[scale=.40]{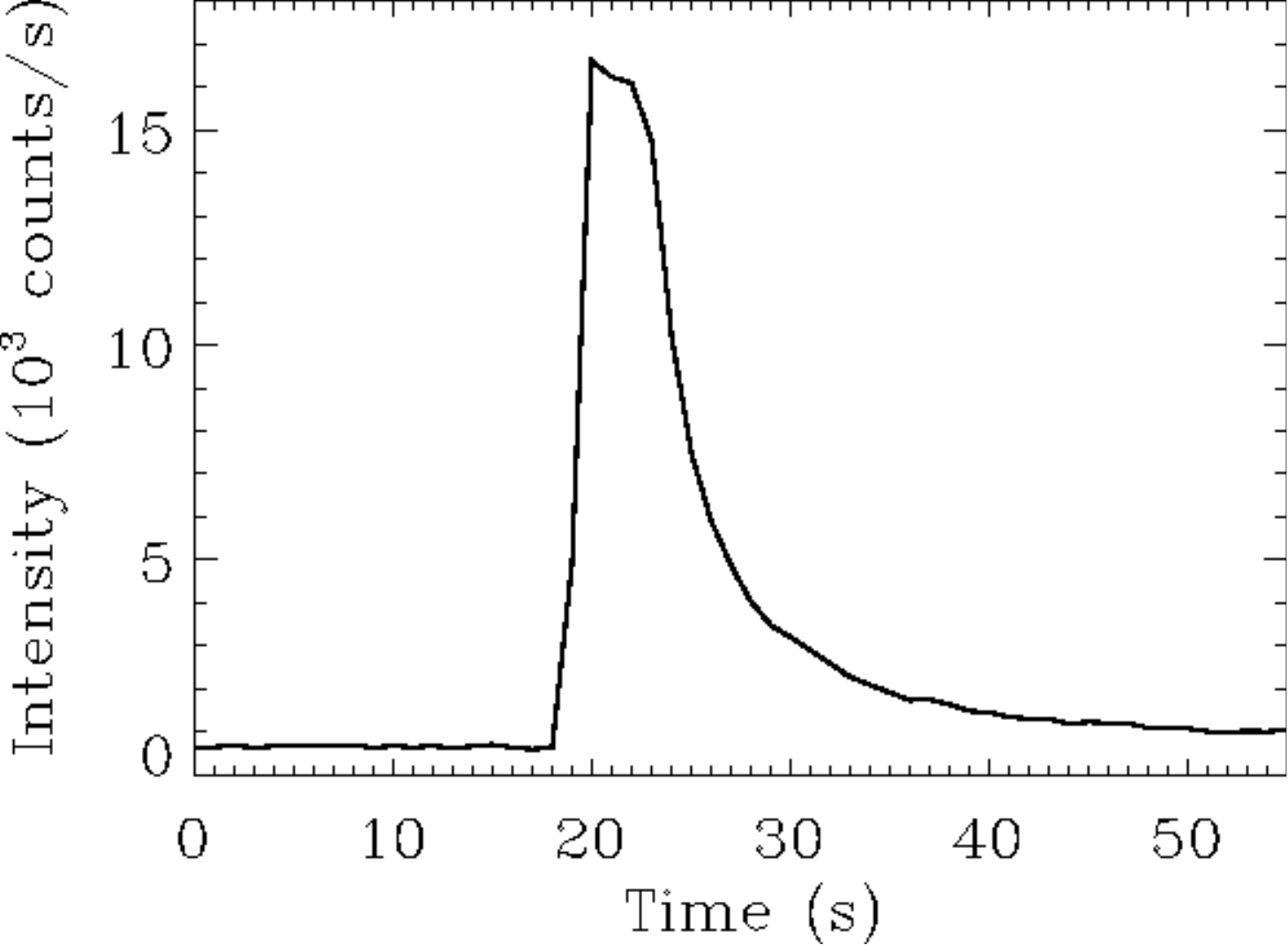}
\caption{The X-ray light curve of a typical thermonuclear (type-I) X-ray burst observed from an LMXB with the {\it Rossi X-ray Timing Explorer} ({\it RXTE}) satellite. The rise, peak and decay parts of the burst are clearly seen (see Sect.~\ref{bursts}).}
\label{burst2}
\end{figure}

Thermonuclear X-ray bursts are eruptions in X-rays, intermittently observed from many neutron star LMXBs \cite{StrohmayerBildsten2006,Galloway2008,Galloway2020}. These bursts were discovered as sharp short-duration intensity rises in X-ray light curves in 1970's \cite{Grindlay1976, Belian1976}.
The most common among these are type-I X-ray bursts, which show an intensity rise typically by a factor of $\sim 10$ in $\approx 0.5-5$~s, a somewhat flat intensity peak, and then a relatively slow decay in $\approx 10-100$~s (see Fig.~\ref{burst2}). For a given LMXB, bursts may occur once in a few hours to days, and the typical energy emitted is $\sim 10^{39}$ erg
\cite{StrohmayerBildsten2006}. Soon after the discovery, it was understood that these bursts originate from the surface of the neutron star. This conclusion was primarily based on the observational fact that the burst emission area, inferred from the energy spectral analysis, matched well with the expected stellar surface area \cite{Swank1977, Hoffmanetal1977a}.
Soon, it was also realized that the nuclear burning of the  accreted matter accumulated on the stellar surface gives rise to these bursts \cite{Joss1977, LambLamb1978, StrohmayerBildsten2006}. The primary evidence for this is, at least for some cases, the ratio of the total fluence of all the bursts to that of the non-burst emission in a data set roughly tallies with the ratio of the expected nuclear energy release per nucleon (a few MeV) to the expected accretion-induced gravitational energy release per nucleon ($\sim 200$ MeV) \cite{Bhattacharyya2010}. This also implies that this nuclear burning is  unstable and intermittent, as otherwise the effect of the nuclear energy release would not be seen as a burst over and above the gravitational energy release due to accretion (see Figure~\ref{burst2}). Such an instability can happen when the nuclear energy generation rate is more temperature sensitive than the cooling rate. However, a stable nuclear burning of the accumulated accreted matter can also happen continuously, which could affect the profile, recurrence time and other properties of thermonuclear bursts.

The observed properties of a burst are determined by its ignition condition, specific nuclear reactions and energy generation, and energy and material transport.
The burst ignition happens in the so-called ocean on an accreting neutron star's crust \cite{Watts2012}. This ocean, made of the accreted matter and its ashes, is expected to have layers of various chemical compositions at various depths. At the top, there should be hydrogen-rich and helium-rich layers of a thickness of hundreds of cm, with a density of $\sim 10^{5-6}$~g~cm$^{-3}$ at the bottom. Below this, there should be first a carbon-rich layer of thickness $\sim 10^4$~cm, and then layers of heavier nuclei \cite{Joss1979,Watts2012}. The thickness and density of layers also depend on previous bursts and stable nuclear burning, as well as the chemical composition of the accreted matter. For example, while hydrogen-rich matter is accreted for many LMXBs, the composition is different for an ultra-compact X-ray binary (UCXB) with a hydrogen deficient and heavier element rich dwarf companion star \cite{zand2008}.
The ignition of the layer of a particular chemical composition depends on the column density and the temperature at the bottom of that layer. Note that the accretion rate is an important parameter which determines when the ignition column density and temperature will be reached. Moreover, the stellar gravity and spin rate, energy injection from the bottom, e.g., from the crust, and a pattern of temperature and chemical distributions, for example, due to the history of previous bursts, could also affect the ignition condition.

The next important aspect of thermonuclear X-ray bursts is nuclear reactions and energy generation, which have the following theoretically identified regimes for increasing accretion rate per unit neutron star surface area (\.{m}) \cite{Bildsten2000}.
{\it Regime 1}: Mixed hydrogen and helium bursts should occur via the CNO (carbon-nitrogen-oxygen) cycle due to an unstable hydrogen ignition for a temperature $T > 10^7$ K and \.{m} $ < 900$ gm cm$^{-2}$ sec$^{-1}$ ($Z_{\rm CNO}/0.01$)$^{1/2}$ ($Z_{\rm CNO}$ is the mass fraction of CNO; \cite{Bildsten2000}).
{\it Regime 2}: For a higher \.{m}, i.e., $900$ gm cm$^{-2}$ sec$^{-1}$ ($Z_{\rm CNO}/0.01$)$^{1/2} < $ \.{m} $< 2\times10^3$ gm cm$^{-2}$ sec$^{-1}$ ($Z_{\rm CNO}/0.01$)$^{13/18}$, which may imply $T > 8\times10^7$ K, hydrogen should burn to produce helium in a stable manner via the ``hot" CNO cycle \cite{Fowler1965}, and hence no thermonuclear burst should be triggered. This, however, is expected to build a helium layer below the hydrogen layer, and if no hydrogen is left when the helium ignition condition is satisfied, we can expect a short ($\sim 10$ s) but very intense helium burst caused by the unstable triple-alpha reaction of the pure helium ($3\alpha$ $\rightarrow$ $^{12}{\rm C}$).
{\it Regime 3}: For an even higher \.{m}, i.e., \.{m} $> 2\times10^3$ gm cm$^{-2}$ sec$^{-1}$
($Z_{\rm CNO}/0.01$)$^{13/18}$,
although hydrogen burns in a stable manner via the ``hot" CNO cycle, sufficient amount of hydrogen can remain unburnt when the helium ignition condition is satisfied. As a result, a mixed hydrogen and helium burst should be triggered by the helium ignition. Such a burst can be much longer ($\sim 100$ s) than a helium burst, and should produce heavy elements beyond the iron group via the rp (rapid-proton) process \cite{Schatz2001}.
{\it Regime 4}: When the accretion rate is very high, i.e, a considerable fraction of the Eddington rate, the nuclear burning should be stable and no burst is expected to occur, as the helium burning becomes less temperature sensitive than the cooling rate \cite{Ayasli1982, Taam1996}.
While different types of bursts, as mentioned above, have been identified from their observed properties, such as duration, peak luminosity and intensity profile, so far it has not been possible to confirm this theoretical picture for most sources, and usually bursts do not occur in a regular manner and in correlation with the accretion rate, as expected. This could be due to the effects of parameters with unknown values, for example the area on which the accreted matter falls, and processes about which we have a limited knowledge, such as the nuclear reaction chain, continuous stable burning and hence the depletion of fuel, ignition conditions as mentioned above, and energy and material transport. Nevertheless, a clear correlation between burst properties and the accretion rate has been observed for the LMXBs GS 1826--238 and IGR J17480--2446 \cite{Gallowayetal2004,ChakrabortyBhattacharyya2011,Chakrabortyetal2011,Linaresetal2012}.

Energy and material transport is the next aspect which decides the observed properties of thermonuclear X-ray bursts. It has been found from the theory and simulations that energy can be transferred by radiation, convection and conduction \cite{Spitkovskyetal2002,Woosleyetal2004,Weinbergetal2006,Maloneetal2011,Watts2012,Garciaetal2018}. Convection causes material transport, and hence is interesting for multiple reasons. For example, convection should contribute to the flame spreading on the neutron star surface after the ignition of a burst \cite{Spitkovskyetal2002}. For photospheric radius expansion (PRE) bursts, i.e., strong bursts for which the radiation pressure temporarily pushes the stellar photosphere away from the neutron star surface, heavy elements could be transported to the stellar surface due to convection, providing a potential tool to probe nuclear reactions and to measure the stellar compactness \cite{Weinbergetal2006}. Convection could also explain the rare thermonuclear bursts with short recurrence times \cite{KeekHeger2017}. This shows that a departure from the above mentioned simple accretion rate based burst model could happen due to complex processes such as convection.

Finally, we note that, in addition to type-I X-ray bursts, long bursts, i.e., bursts of longer durations of $\approx 30$~min and
fluence of $\approx 10^{41}$~erg have been observed from several
LMXBs (e.g., \cite{Kapteinetal2000, intZandetal2002, intZandetal2005}). 
Even longer ($\sim 1-3$ hours) and more energetic ($\sim 10^{42}$ erg) thermonuclear bursts, viz., superbursts, have also been observed from some sources (e.g., \cite{Cornelisse2000, Kuulkers2004,
intZandetal2004, Kuulkers2005, Keek2008}). The latter bursts could originate from the thermally unstable $^{12}{\rm C}$ fusion in the carbon-rich layer of the ocean \cite{WoosleyTaam1976,
TaamPicklum1978, BrownBildsten1998}.

\subsection{Burst oscillations: discovery and growth of the field}\label{BO}

With this basic knowledge of thermonuclear X-ray bursts, here we will briefly describe how the field of nuclear-powered pulsations, or burst oscillations, has developed over the last two-and-half decades. As soon as the first MSP was discovered in radio \cite{bac82}, it was proposed that an MSP is an old neutron star spun-up in its LMXB phase, in which it gains angular momentum by accretion \cite{rad82,alp82}.
But no millisecond period in an LMXB was found for more than a decade. In 1995, the {\it Rossi X-ray Timing Explorer} ({\it RXTE}) satellite was launched. The Proportional Counter Array (PCA) on-board {\it RXTE} had a large area and high time resolution, and hence was ideal to detect millisecond period X-ray signals \cite{Jahodaetal2006}. Indeed, soon after the launch, coherent millisecond period brightness oscillations at a frequency of $\approx 363$ Hz were discovered in six thermonuclear X-ray bursts observed from the LMXB 4U 1728--34 \cite{Strohmayeretal1996}. The root mean square (rms) amplitudes of this timing feature were in the range of  2.5\%--10\%, and the frequency had an upward drift of $\approx 1.5$ Hz for a few seconds, after which the oscillations became almost coherent during the burst decay. Because of this effective coherence, and since a near-millisecond spin period is  expected for a neutron star in an LMXB, the burst oscillation frequency was thought to be the stellar spin frequency. Therefore, the simplest explanation of this feature was a spin modulated inhomogeneous burst emission, which could explain the observed rms amplitudes.

Burst oscillations from multiple neutron star LMXBs were discovered in the next few years \cite{Strohmayeretal1999,Strohmayeretal2001}, and the finding of a remarkable stability of the asymptotic oscillation frequency of two sources over a period of $\sim 1.5$ years strongly indicated that this frequency is the neutron star spin frequency \cite{Strohmayeretal1998}.
Nevertheless, this could not be established beyond doubt. In fact, based on a model of another ~millisecond period timing feature, viz., kilohertz quasiperiodic oscillations \cite{vanderKlis2006}, as well as some observational indications, it was proposed that the burst oscillation frequency could be twice the stellar spin frequency \cite{Miller1999}. But in 2003, Chakrabarty et al. reported that the oscillation frequency in the decay phase of thermonuclear bursts from the AMXP SAX J1808.4--3658 matched well with the known spin frequency ($\approx 401$~Hz) of this source \cite{Chakrabartyetal2003}. This not only showed that a burst oscillation frequency is the neutron star spin frequency, but also provided a new way to measure the latter frequency.
Detection of burst oscillations from more AMXPs has confirmed this (see Table~\ref{table1}).

\begin{figure}[t]
\sidecaption[t]
\includegraphics[scale=.40]{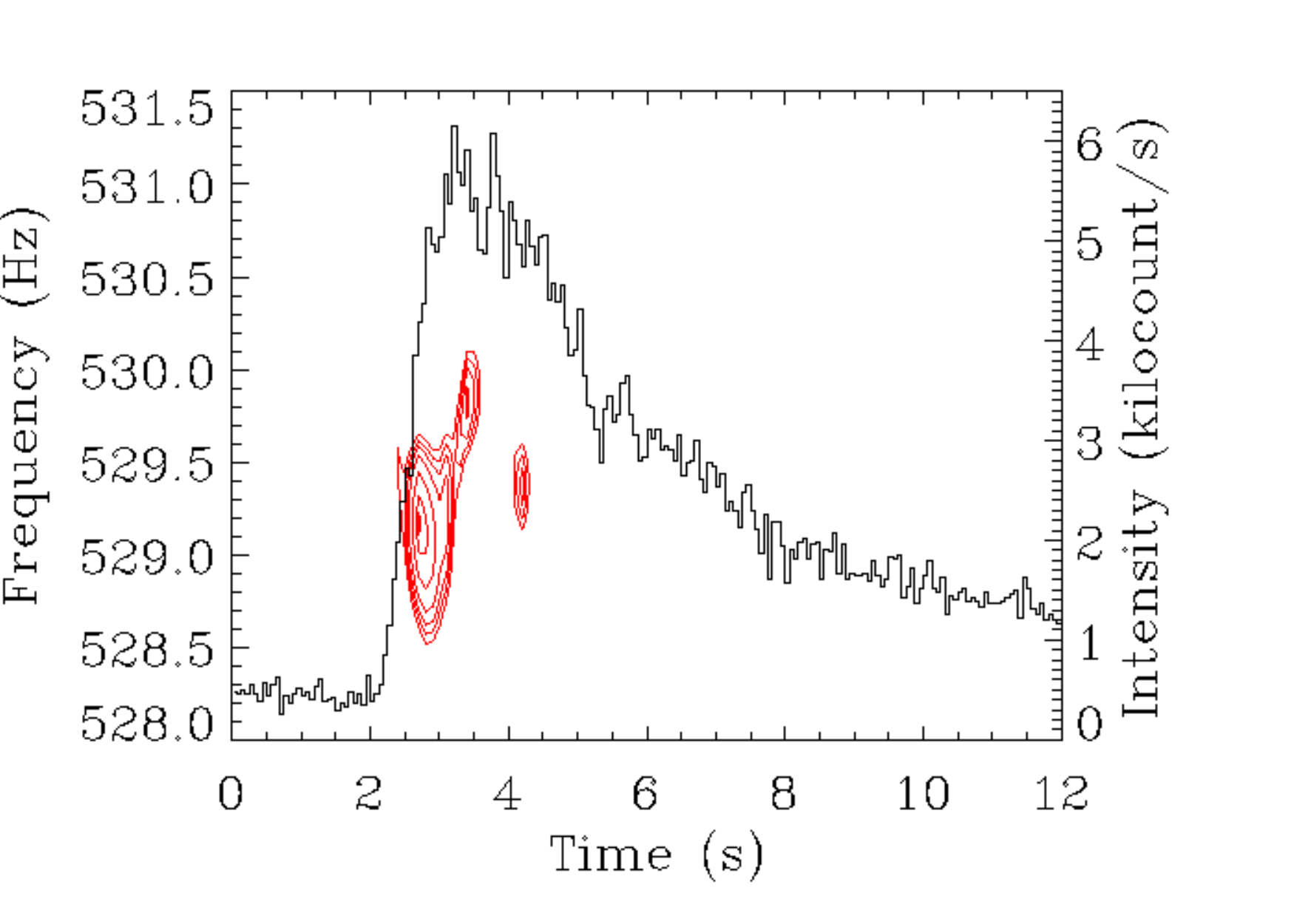}
\caption{A thermonuclear X-ray burst observed from the neutron star LMXB 1A 1744--361 with the {\it Rossi X-ray Timing Explorer} ({\it RXTE}) satellite \cite{Bhattacharyyaetal2006}. Dynamic power spectrum, i.e., X-ray brightness variation power as a function of frequency and time, is shown. The corresponding power contours (in red) suggest that burst oscillations can be present during only portions of a given burst, and their frequency could evolve (see Sect.~\ref{BO}).}
\label{bo}
\end{figure}

The firm identification of the burst oscillation frequency with the neutron star spin frequency confirmed a new type of pulsars -- the NMXPs. This also made the burst oscillation feature a promising tool to study the strong gravity and to measure some parameter values of neutron stars \cite{Bhattacharyyaetal2005, Bhattacharyya2010}. The latter is required to probe the super-dense degenerate core matter of these stars (see 
Chap.~9).

\begin{table}[hbtp]
\caption{Nuclear-powered millisecond X-ray pulsars.}
\scriptsize
\label{table1}
\begin{tabular}{p{0.5cm}p{3.0cm}p{1.5cm}p{6.3cm}}
\hline\noalign{\smallskip}
No. & Source name & Frequency & Remarks  \\
 &  & (Hz) & \\
\noalign{\smallskip}\hline\noalign{\smallskip}
1 & 4U 1608--522$^{\rm T}$ & 620 & Oscillations in multiple bursts were reported using {\it RXTE} PCA observations  \cite{Hartmanetal2003,Galloway2008,Watts2012}. \\
2 & SAX J1750.8--2900$^{\rm T}$ & 601 & A $5\sigma$ detection of oscillations in a burst was reported in 2002 \cite{Kaaretetal2002} using {\it RXTE} PCA observations. Later, oscillations in more bursts were reported \cite{Galloway2008}. \\ 
3 & GRS 1741.9--2853$^{\rm T}$ & 589 & Oscillations in multiple bursts were reported using {\it RXTE} PCA observations \cite{Strohmayeretal1997,Galloway2008}. \\
4 & 4U 1636--536 & 582 & Oscillations in many bursts were reported using {\it RXTE} PCA observations \cite{Strohmayeretal1998a,Gilesetal2002,StrohmayerMarkwardt2002,Galloway2008}. \\
5 & MXB 1659--298$^{\rm T}$ & 567 & Oscillations in multiple bursts were reported using {\it RXTE} PCA observations \cite{Wijnandsetal2001,Galloway2008}. \\
6 & EXO 0748--676$^{\rm T}$ & 552 & Oscillations in two bursts were reported in 2010 using {\it RXTE} PCA observations with a significance of $6.3\sigma$ \cite{Galloway2010}. \\
7 & Aql X--1$^{\rm T, IP}$ & 550 & Oscillations in multiple bursts were reported using {\it RXTE} PCA observations \cite{Zhangetal1998,Watts2012}. \\
8 & SAX J1810.8--2609$^{\rm T}$ & 532 & A $5.75\sigma$ detection of oscillations in one burst was reported in 2018 using {\it RXTE} PCA observations \cite{Bilousetal2018}. \\
9 & 1A 1744--361$^{\rm T}$ & 530 & Oscillations in one burst were reported with a significance of $5.02\times10^{-7}$ in 2006 using {\it RXTE} PCA observations \cite{Bhattacharyyaetal2006}. Later, oscillations in the same burst were confirmed \cite{Galloway2008}. \\
10 & KS 1731--260$^{\rm T}$ & 524 & Oscillations in multiple bursts were reported using {\it RXTE} PCA observations \cite{Smithetal1997,Galloway2008}. \\
11 & SAX J1808.4--3658$^{\rm T, P}$ & 401 & Oscillations, with frequencies close to the accretion-powered pulsation frequency, in multiple bursts were reported using {\it RXTE} PCA observations \cite{Chakrabartyetal2003,Galloway2008}. \\
12 & IGR J17498--2921$^{\rm T, P}$ & 401 & Oscillations, with frequencies close to the accretion-powered pulsation frequency, in two bursts were reported using {\it RXTE} PCA observations, one with $4.1\sigma$ significance  \cite{Linaresetal2011,ChakrabortyBhattacharyya2012} and another with $8.8\sigma$ significance \cite{ChakrabortyBhattacharyya2012}.\\
13 & HETE J1900.1--2455$^{\rm T, IP}$ & 377 & Oscillations in one burst were reported in 2009 using {\it RXTE} PCA observations \cite{Wattsetal2009}. While the oscillations are relatively weak, the feature is considered robust due to the closeness of the corresponding frequency to the known accretion-powered pulsation frequency \cite{BilousWatts2019}.  \\
14 & 4U 1728--34 & 363 & Oscillations in many bursts were reported using {\it RXTE} PCA observations \cite{Strohmayeretal1996,Galloway2008}.\\
15 & 4U 1702--429 & 329 & Oscillations in many bursts were reported using {\it RXTE} PCA observations \cite{Markwardtetal1999,Galloway2008}. \\
16 & XTE J1814--338$^{\rm T, P}$ & 314 & Oscillations, with frequencies close to the accretion-powered pulsation frequency, in many bursts were reported using {\it RXTE} PCA observations \cite{Strohmayeretal2003,Galloway2008}. \\
17 & IGR J17191--2821$^{\rm T}$ & 294 & Oscillations in multiple bursts were reported using {\it RXTE} PCA observations \cite{Altamiranoetal2010}. \\
18 & IGR J18245--2452$^{\rm T, tP}$ & 254 & Oscillations, with a frequency close to the accretion-powered and spin-powered pulsation frequencies, in one burst were reported in 2013 using {\it Swift} XRT observations \cite{Patruno2013,pap13}. \\
19 & IGR J17511--3057$^{\rm T, P}$ & 245 & Oscillations, with frequencies close to the accretion-powered pulsation frequency, in multiple bursts were reported using {\it RXTE} PCA observations \cite{Altamiranoetal2010a}.\\
\noalign{\smallskip}\hline\noalign{\smallskip}
\end{tabular}
$^{\rm T}$ Transient source. $^{\rm P}$ AMXP. $^{\rm IP}$ Intermittent AMXP. $^{\rm tP}$ Transitional MSP.
\end{table}

Currently, 19 NMXPs are known (see Table~\ref{table1}), implying about 20\% of burst sources are NMXPs \cite{Galloway2020}. In addition, tentative detection of burst oscillations has been reported for six neutron star LMXBs (see Table~\ref{table2}). Note that we have not included the burst oscillation source IGR J17480--2446, because the neutron star in this LMXB has a spin frequency of 11 Hz \cite{StrohmayerMarkwardt2010}, and hence is not an MSP. Detection and observational studies of burst oscillations from so many sources, efforts to theoretically model and understand them, and use of this timing feature as a tool to probe various source properties have firmly established the field of NMXPs over the last two-and-half decades.

Nevertheless, it is quite challenging to detect burst oscillations and study their properties for the following reasons. Firstly, they appear only during thermonuclear bursts, which are typically irregular and have short duration and low duty cycle. In addition, most of the known NMXPs are transient sources (see Table~\ref{table1}), and we usually observe thermonuclear bursts during their outbursts, which are irregular and of low duty cycle. Next, these oscillations are usually seen only for a fraction of bursts even for known NMXPs, and even for a given burst this feature may be seen intermittently and for a small fraction of time, and with a frequency evolution (see Fig. ~\ref{bo}). This is why the detection of burst oscillations often crucially depends on the used criteria \cite{Watts2012,Bilousetal2018,Galloway2020}. Finally, not all X-ray instruments can detect this high-frequency feature. In fact, burst oscillations from all sources, except two, have so far been discovered or reported using {\it RXTE} PCA (see Tables~\ref{table1} and \ref{table2}). After {\it RXTE} was decommissioned in 2012, now {\it AstroSat} and {\it Neutron Star Interior Composition Explorer} ({\it NICER}) satellites are ideal to detect burst oscillations \cite{VerdhanChauhanetal2017,Mahmoodifaretal2019,Bultetal2019}.

\begin{table}[hbtp]
\caption{Neutron star LMXBs with tentative detection of burst oscillations.}
\scriptsize
\label{table2}
\begin{tabular}{p{0.5cm}p{3.0cm}p{1.5cm}p{6.3cm}}
\hline\noalign{\smallskip}
No. & Source name & Frequency & Remarks  \\
 &  & (Hz) & \\
\noalign{\smallskip}\hline\noalign{\smallskip}
1 & XTE J1739--285$^{\rm T}$ & 1122/386.5 & Oscillations at 1122 Hz with $\lsim 4\sigma$ significance in one burst were reported in 2007 using {\it RXTE} PCA observations \cite{Kaaretetal2007}. However, this detection was not confirmed later for the same burst for independent time windows \cite{Galloway2008}. Another tentative detection of oscillations at 386.5 Hz was reported in 2021 from a burst observed with {\it NICER} \cite{Bultetal2021}.  \\
2 & GS 1826--24$^{\rm T}$ & 611 & Burst oscillations with $< 4\sigma$ significance were reported in 2005 using {\it RXTE} PCA observations \cite{Thompsonetal2005}. \\
3 & 4U 0614+09 & 415 & Oscillations with $\sim 4\sigma$ significance in one burst were reported in 2008 using {\it Swift} BAT observations \cite{Strohmayeretal2008}. \\
4 & MXB 1730--335$^{\rm T}$ & 306 & Burst oscillations with $\sim 2.5\sigma$ significance were reported in 2001 using {\it RXTE} PCA observations \cite{Foxetal2001}. \\
5 & 4U 1916--053 & 270 & Oscillations with $\sim 4.6\sigma$ significance in one burst were reported in 2001 using {\it RXTE} PCA observations \cite{Gallowayetal2001}. \\
6 & XB 1254--690 & 95 & A $2\sigma$ candidate oscillation signal in one burst was reported in 2007 using {\it RXTE} PCA observations \cite{Bhattacharyya2007}. \\
\noalign{\smallskip}\hline\noalign{\smallskip}
\end{tabular}
$^{\rm T}$ Transient source.
\end{table}

We will discuss the main observational aspects of burst oscillations in Sect.~\ref{Observational}, except some of those related to thermonuclear flame-spreading on neutron stars.
The theory and observations of such spreading, as well as the origin of oscillations during burst rise, will be presented in Sect.~\ref{flame-spreading}. In Sect.~\ref{Theory}, we will discuss what may cause oscillations during burst decay. Finally, we will mention why burst oscillations are important and what is required to use them as a reliable tool, in Sect.~\ref{Conclusion}.

\section{Observational aspects}\label{Observational}

In this section, we will discuss the main observational properties of burst oscillations. While burst oscillation models will be described in subsequent sections, here we will briefly mention certain physical effects, which might explain some of the observational aspects.

\subsection{Frequency and coherence}\label{Frequency}

Burst oscillation frequency drifts by $\sim 1-3$ Hz in many thermonuclear X-ray bursts, but remains within a few Hz of the known neutron star spin frequency for AMXPs. A further evidence of the  stellar spin connection of burst oscillation frequencies comes from a transitional MSP IGR J18245--2452 
\cite{pap13}, because it has manifested itself as each of  spin-powered, accretion-powered and nuclear-powered MSPs (Table~\ref{table1}). However, somewhat significant burst oscillation signals, at frequencies far from the known or believed stellar spin frequencies, have also been reported for some sources. For example, a $\sim 5.5\sigma$ oscillation signal was found at 45 Hz in the average power spectrum of many thermonuclear bursts from EXO 0748--676 in 2004 \cite{VillarrealStrohmayer2004}. But in 2010, $6.3\sigma$ 
oscillations in two bursts from the same source were reported at 552 Hz \cite{Galloway2010}, which is believed to be approximately the neutron star spin frequency. Note that the 45 Hz signal was not confirmed by further analyses \cite{BilousWatts2019}.
Similarly, an oscillation signal at 410 Hz was reported for a burst from SAX J1748.9--2021 \cite{Kaaretetal2003}. But this source was later found to be an intermittent AMXP with the stellar spin frequency of 442 Hz, and the significance of the 410 Hz signal decreased below $3\sigma$ for a revised number of trials \cite{Altamiranoetal2008}.

Burst oscillation frequency typically drifts upwards, although exceptions, such as a downward drift and the presence of two simultaneous oscillations at frequencies separated by a few Hz, are observed in $\sim 5$\% cases \cite{Munoetal2002}. Upward drifting frequencies usually approach an apparent saturation frequency, known as the asymptotic frequency $\nu_{\rm a}$. This frequency is determined by modelling the frequency evolution with the formula $\nu(t) = \nu_{\rm a}(1-\delta_{\nu}e^{-t/\tau})$, where $t$, $\delta_{\nu}$, and $\tau$ are time, and fractional drift and drift timescale of the frequency, respectively \cite{StrohmayerBildsten2006}. Note that the frequency evolution is usually estimated using the phase connection technique \cite{Munoetal2000,Munoetal2002}. The fractional dispersion in asymptotic frequencies over several years has been found to be typically less than $10^{-3}$ for multiple sources (\cite{Munoetal2002,StrohmayerBildsten2006,Watts2012} and references therein). Such a high stability strongly points to the neutron star spin frequency, and hence $\nu_{\rm a}$, when measured, is usually considered to be the stellar spin frequency for non-AMXPs. This assumption has been tested for two intermittent AMXPs, Aql X--1 and HETE J1900.1--2455, for which $\nu_{\rm a}$ is 0.5 Hz and $\sim 1$ Hz below the known stellar spin frequencies, respectively (\cite{Watts2012} and references therein). 

\begin{figure}[t]
\sidecaption[t]
\includegraphics[scale=.50]{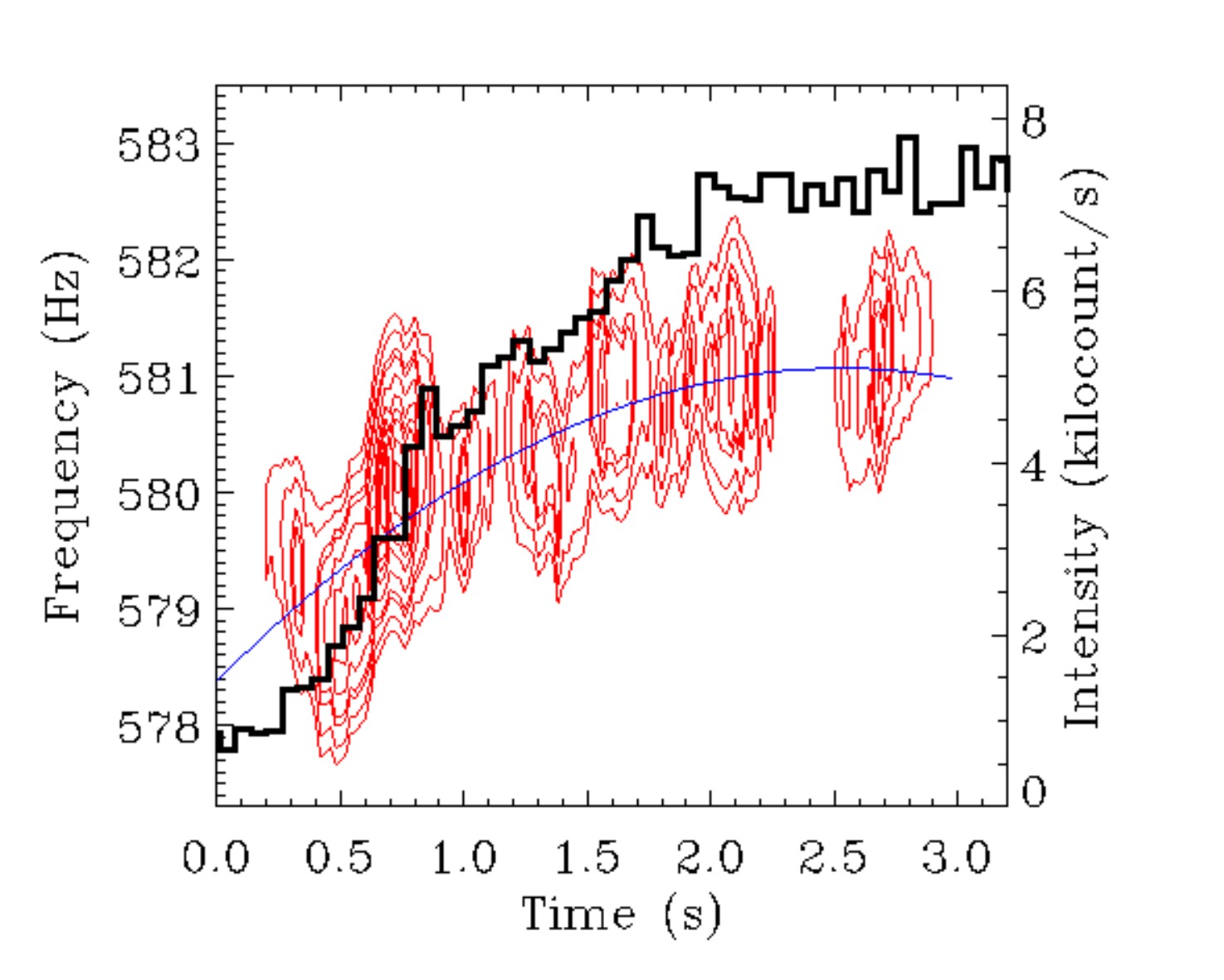}
\caption{Rise of a thermonuclear X-ray burst (black histogram) observed from the neutron star LMXB 4U 1636--536 with the {\it Rossi X-ray Timing Explorer} ({\it RXTE}) satellite \cite{BhattacharyyaStrohmayer2005}. Dynamic power spectrum, i.e., burst rise oscillation power as a function of frequency and time, is shown. The corresponding power contours (in red) and the best-fit frequency evolution curve (in blue) suggest that the  burst oscillation frequency drifts upward by a few Hz to an apparent saturation value (see Sect.~\ref{Frequency}).}
\label{1636_bo}
\end{figure}

The frequency evolution $\nu(t)$ is characterized either by fitting data with a formula, such as $\nu(t) = \nu_{\rm a}(1-\delta_{\nu}e^{-t/\tau})$, $\nu(t) = \nu_{\rm 0} + \dot\nu t + \ddot\nu t^2$, etc., for the entire oscillation duration, or using several constant frequency values in shorter time bins. The modelling with a formula can give a highly coherent signal, with a coherence $Q (= \nu/\Delta\nu$) as high as, for example, $\sim 4000$ \cite{StrohmayerMarkwardt1999}.
Note that an evolution of frequency is not detected for many burst oscillation signals for the following reasons: oscillations are weak, short-lived or intermittent, or genuinely there is no evolution. Even when there is a frequency drift, the evolution may not be smooth on short timescales. 
This may be because of  two simultaneous signals with a small frequency difference, or a discontinuous frequency evolution, or discrete phase jumps \cite{Munoetal2002}.
The saturation or asymptotic frequency is usually measured in the decay phase, because a burst rise duration is typically only $\sim 1$~s, and hence it is more challenging to detect a frequency evolution in the rising phase. Nevertheless, such an evolution and an indication of a saturation frequency have been found for the rise of some bursts, a clear example of which is shown in Fig.~\ref{1636_bo}.

A drift of a burst oscillation frequency, typically $\approx 1$\% in a few seconds, cannot be caused by the stellar spin evolution, as that would require an unphysically large torque. 
Note that an azimuthally asymmetric brightness pattern on the spinning neutron star surface should give rise to burst oscillations.
Therefore, in order to explain the frequency drift, this pattern may move and/or change shape/size in the azimuthal direction. A simple model proposed to explain this is the following \cite{Strohmayeretal1997}. A $\sim 10$ m thick layer of accumulated accreted matter could expand to a thickness of about $\sim 30$ m during a burst, and hence in order to conserve the angular momentum, its rotation frequency should decrease by $\delta\nu \approx 2\nu (20 {\rm m}/R)$, where $\nu$ and $R$ are the stellar spin frequency and radius, respectively \cite{StrohmayerBildsten2006}. This may explain why at the beginning we observe a burst oscillation frequency, which is typically $\sim 2$~Hz less than the stellar spin frequency. As the burning layer cools down, its thickness decreases, and hence the burst oscillation frequency is expected to increase towards the stellar spin frequency in a few seconds, as observed. However, this explanation has following problems. (1) Observations of a frequency drift of $\approx 3.6$~Hz for 4U 1916--053 \cite{Gallowayetal2001} and $\sim 5$~Hz for MXB 1659--298 \cite{Wijnandsetal2001} may imply an unphysically large increase of burning layer thickness during the burst. (2) A frequency increase by a few Hz has also been observed during the rising phase of bursts (see Fig.~\ref{1636_bo} for an example), when the burning layer should not cool down. (3) If the burning layer slips at a rate of a few Hz with respect to the neutron star, an azimuthal asymmetry could disappear in a fraction of a second, and hence burst oscillations might not be observed for as long as a few seconds.

\subsection{Amplitude}\label{Amplitude}

The average fractional root mean square (rms) amplitude of burst oscillations is typically in the range of $0.02-0.2$, with a median of $\sim 0.05$ \cite{Watts2012,Galloway2008,Munoetal2002a}. However, for burst oscillations detected at the beginning of the rise of bursts, the fractional rms amplitude is typically found to be higher ($\gsim 0.2$, and sometimes as high as $\sim 0.6$; \cite{ChakrabortyBhattacharyya2014,Strohmayeretal1998a}). It was also noticed soon after the discovery of the burst oscillation feature \cite{Strohmayeretal1997a,Strohmayeretal1998a}, and later confirmed with extensive analyses \cite{ChakrabortyBhattacharyya2014}, that the burst oscillation amplitude decreases with time during burst rise. This is expected for thermonuclear flame-spreading on the neutron star surface, as we will discuss in Sect.~\ref{flame-spreading}. 

If oscillations are detected during both rise and decay portions of a burst, the oscillation amplitude can decrease until the burst peak, and then can again increase and finally decrease in the decay phase \cite{Mahmoodifaretal2016}. The fractional rms amplitude during burst decay is typically small ($\sim 0.1$; \cite{Galloway2008,Mahmoodifaretal2019}). But recently very large burst decay oscillation amplitudes (fractional rms $\sim 0.4-0.5$) have been measured for two bursts observed from 4U 1728--34 with the {\it NICER} satellite \cite{Mahmoodifaretal2019}. These oscillations, however, have been detected above 6 keV, and hence are somewhat unusual. Such high amplitude values could be very useful to constrain models (see Sect.~\ref{Theory}).

Burst oscillation amplitude usually changes throughout a burst, sometimes smoothly, but in other cases erratically.
While oscillations can be detected in any portion of a burst -- rise, peak or decay, they do not appear during the peak of a PRE burst, confirming that oscillations are stellar surface phenomena.
The short durations and/or intermittence of oscillations in many bursts (see Sect.~\ref{Frequency}) could be due to low observed X-ray count rates and/or apparently erratic low oscillation amplitudes. From a systematic study of the rise of 161 bursts from 4U 1636--536, it was found that intermittent non-detections of oscillations cannot be solely due to low observed X-ray count rates \cite{ChakrabortyBhattacharyya2014}. Detection of burst decay oscillations, on the other hand, could be linked to the apparent burst emission area evolution
estimated from spectral analyses, as found for two NMXPs -- 4U 1636--53 and 4U 1728--34 \cite{Zhangetal2013,Zhangetal2016}.

It has also been found for non-AMXPs that the oscillation fractional amplitude of the strongest signal during a burst can be both low and high ($\sim 0.05-0.2$) in relatively softer source spectral states, but is typically low ($\sim 0.05-0.1$) in harder spectral states \cite{Galloway2020,Ootesetal2017}. This may cause  burst oscillations to be usually detected in softer states, which are thought to be linked with higher accretion rates \cite{Munoetal2004,Galloway2008,Galloway2020}. Oscillations are also mostly found for bursts with higher peak fluxes and shorter durations, which are likely helium bursts on many occasions (see Sect.~\ref{bursts}), and are often PRE bursts \cite{Galloway2020}. However, detection of oscillations does not depend on separation time and fluence of bursts \cite{Galloway2020}.

\subsection{Harmonic content}\label{Harmonics}

Burst oscillations could originate from a hot spot or bright region on a spinning neutron star surface, and Doppler
and time delay effects are expected to make the oscillation profile more asymmetric and narrowly peaked (\cite{Munoetal2002a} and references therein). This would cause larger amplitudes of the fundamental and overtones of oscillations. In such a case, the first overtone of the signal is expected to be detected, at least for some bursts. But a harmonic content has not been significantly detected for any individual burst from a non-AMXP or an intermittent AMXP \cite{Munoetal2002a,Wattsetal2009}. A theoretical study suggested that such a lack of harmonic content implies a hot spot near the spinning pole or a hot spot covering nearly half the stellar surface \cite{Munoetal2002a}. It is not known if one of these could be true for burst decay oscillations, since the cause of such oscillations is not yet understood (Sect.~\ref{Theory}). However, during the burst rise, an expanding hot spot due to the thermonuclear flame-spreading is believed to cause oscillations (see Sect.~\ref{flame-spreading}). 
In this case, while a large hot spot is expected towards the end of the burst rise, the hot spot should be relatively small at the beginning (Sect.~\ref{flame-spreading}). Hence, unless the spot is very close to the spinning pole, there should be a significant harmonic content in oscillations at the beginning of a burst. Indeed, a significant first overtone was found at the beginning of burst rise, by stacking several bursts observed from the non-AMXP 4U 1636--536 with {\it RXTE} \cite{BhattacharyyaStrohmayer2005}. A stacking was required to improve the statistics, as the observed X-ray count rates were low at the beginning of burst rise.

\subsection{Energy dependence}\label{Energydependence}

Burst oscillation fractional amplitude increases with photon energy for non-AMXPs and intermittent AMXPs \cite{Munoetal2003,Wattsetal2009}, although such a trend is not always observationally clear \cite{Chakrabortyetal2017}. This trend is not unexpected, as the non-burst emission of a source peaks in a lower energy band, and hence the brightness contrast between burst and non-burst emissions are lesser in lower energies. Such an increasing brightness contrast with energy should cause an increasing fractional amplitude, as observed \cite{Chakrabortyetal2017}.

Photons of one energy band may arrive later than those of another band. If higher energy photons follow the lower energy photons, then it is a hard lag, and if the former photons precede the latter ones, then it is a soft lag. Detection of such an energy dependent phase lag for burst oscillations could be very useful to probe the origin of this timing feature. For example, a hard lag may imply that softer photons from a burst are upscattered to higher energy photons by a hot electron cloud \cite{LeeMiller1998}. A soft lag, on the other hand, could be produced by one of the following mechanisms. (1) As the hot spot on a spinning stellar surface moves towards the observer, the photons emitted from this spot at an earlier spin phase should be Doppler-shifted to higher energies \cite{Ford1999}. (2) High energy photons could be Compton downscattered in a relatively cool atmosphere \cite{Cuietal1998}. (3) A hot spot could laterally expand and cool \cite{Cuietal1998}. (4) Burst photons could be scattered by an accretion disk rotating in the same direction of the neutron star spin \cite{SazonovSunyaev2001}.

An indication of hard lags for burst decay oscillations was found by averaging groups of bursts observed from individual sources \cite{Munoetal2003}. However, it was later reported that the phase lags of oscillations during some of these individual bursts were consistent with soft lags \cite{Artigueetal2013}. More recently, a detailed analysis of {\it RXTE} data from several sources has shown that oscillations from different bursts from the same source could have either of no lag, hard lag, soft lag or mixed lag \cite{Chakrabortyetal2017}. Explanation of a hard lag is challenging, because a Compton upscattering should decrease the fractional amplitude of oscillations, while relatively large fractional amplitudes in high-energy bands have been observed for hard lags \cite{Munoetal2003}. In addition, it has recently been found for the soft state of the neutron star LMXB 4U 1728--34 that the burst emission is not significantly reprocessed by a  hot electron cloud \cite{Bhattacharyyaetal2018}. Moreover, it is particularly challenging to explain the hard lag for oscillations during the burst rise \cite{Chakrabortyetal2017}, when an expanding hot spot on the spinning neutron star might cause a soft lag due to the Doppler shift \cite{Ford1999}. However, the current detection of phase lags are mostly marginal \cite{Watts2012}, and a stronger detection with future instruments will be required to understand burst oscillation phase lags.

\subsection{Connection with accretion-powered pulsations}\label{AMXP}

Similar to burst oscillations, accretion-powered pulsations also originate from an azimuthal asymmetry on or near the neutron star surface. For the latter feature, a sufficiently strong stellar magnetic field channels the accreted matter on to magnetic poles
(see 
Chap.~4).
The resulting hot polar regions, which are rotating hot spots, can give rise to the observed  accretion-powered pulsations. Studies of those sources, which are both NMXPs and AMXPs, particularly persistent AMXPs, can therefore be useful to probe several important questions, such as, if stellar magnetic field and polar regions contribute to the burst oscillation mechanism and if this mechanism is different for AMXPs and non-AMXPs  in some respects. To this end, the first aspect to check is how the burst oscillation frequency behaves with respect to the  accretion-powered pulsation frequency, or the {\it pulsar frequency}. Oscillation frequency does not significantly drift during the burst decay of persistently accretion-powered and nuclear-powered X-ray millisecond pulsars (ANMXPs), and typically remains close to the pulsar frequency. While for XTE J1814--338, the burst oscillation frequency is extremely stable, being within $\sim 10^{-8}$~Hz of the pulsar frequency, for SAX J1808.4--3658, IGR J17511--3057 and IGR J17498--2921, the former frequency is within a few mHz, $\sim 0.05$~Hz and $\sim 0.25$~Hz of the latter frequency, respectively \cite{Chakrabartyetal2003,Altamiranoetal2010a,Strohmayeretal2003,Wattsetal2005,Wattsetal2008,ChakrabortyBhattacharyya2012,Watts2012}.
These indicate that during the burst decay of ANMXPs, and particularly for XTE J1814--338, burst oscillation and accretion-powered pulsation mechanisms are closely related.

\begin{figure}[t]
\sidecaption
\includegraphics[scale=.50]{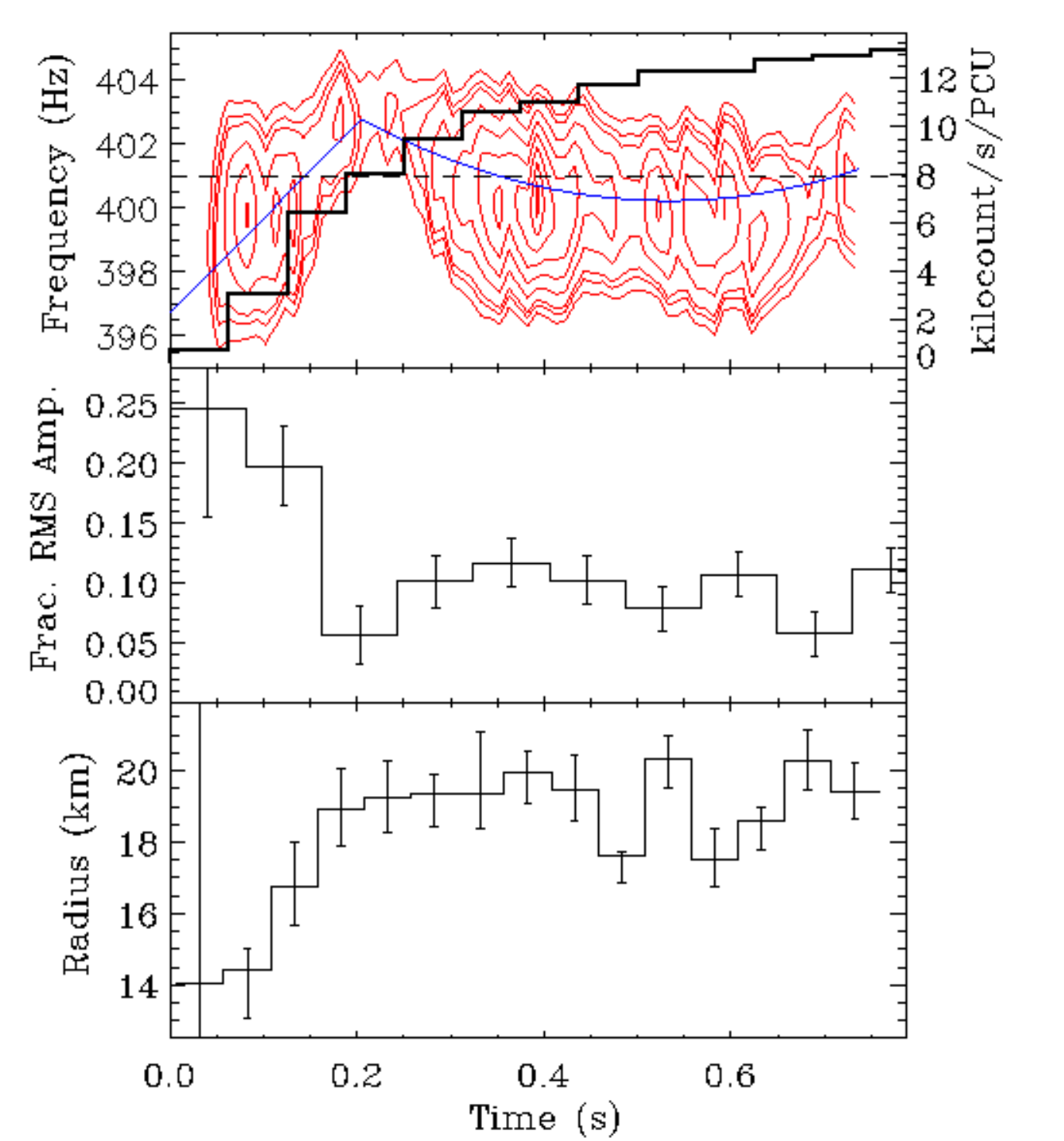}
\caption{{\it Upper panel}: Rise of a thermonuclear X-ray burst (black histogram) observed from the AMXP  SAX J1808.4--3658 with the {\it Rossi X-ray Timing Explorer} ({\it RXTE}) satellite. 
The power contours (in red) of the dynamic power spectrum of burst rise oscillations, the best-fit frequency evolution curve (in blue) and the pulsar spin frequency (the horizontal dashed line) are shown. {\it Middle panel}: The fractional rms amplitude evolution of burst rise oscillations. {\it Lower panel}: Evolution of an apparent radius (assuming 10 kpc source distance), which is proportional to the square root of the burning area, as inferred from spectral fitting (\cite{BhattacharyyaStrohmayer2006}; see Sects.~\ref{AMXP} and \ref{Further}).}
\label{1808_bo}
\end{figure}

For the burst rise of SAX J1808.4--3658, however, the burst oscillation frequency increased from a lower value by a few Hz, and appeared to overshoot the pulsar frequency \cite{Chakrabartyetal2003,BhattacharyyaStrohmayer2006} (see also Fig.~\ref{1808_bo}). This and other simultaneous burst rise properties indicated an expanding hot spot due to thermonuclear flame-spreading \cite{BhattacharyyaStrohmayer2006}, which should be unrelated to accretion-powered pulsations. An upward frequency drift of $\sim 0.1$~Hz and a similar overshooting was also reported for IGR J17511--3057 \cite{Altamiranoetal2010a,Watts2012}. But for XTE J1814--338, the oscillation frequency did not drift even during the burst rise, except for one burst, for which the frequency decreased by $\sim 0.1$~Hz.

The next aspect to check is if burst oscillations and accretion-powered pulsations of an ANMXP are phase-locked when their frequencies are consistent with each other, and if so, what the constant phase offset is. A phase-locking with the zero phase offset might indicate that almost the same hot spot, that is the one at a magnetic polar cap, generates the two sets of pulsations. One implication of a non-zero offset might be burst oscillations due to a hot spot located at a constant longitude separation with a polar cap. However, burst oscillations could have soft or hard lags (see Sect.~\ref{Energydependence}) and accretion-powered pulsations typically have soft lags \cite{Cuietal1998,Watts2012}. Therefore, the phase offset between these two sets of pulsations can also be energy dependent. This was found for the ANMXP XTE J1814--338. For this source, while the two sets of pulsations were phase-locked with a zero offset at lower energies, there could be a non-zero offset at higher energies \cite{Strohmayeretal2003,Wattsetal2008}. This is because, while accretion-powered pulsations showed soft lags of up to 50 ${\rm \mu s}$, no evidence of lags was found for burst oscillations \cite{WattsStrohmayer2006}. Accretion-powered pulsations and burst oscillations have been reported to be phase-locked also for another ANMXP, IGR J17511--3057 \cite{Altamiranoetal2010a,Watts2012}.

Burst oscillations appear to be present in all observed bursts of ANMXPs. Moreover, oscillations were observed throughout the bursts, except during the peak of PRE bursts, for two ANMXPs: SAX J1808.4--3658 and XTE J1814--338. These aspects are different from other burst oscillation sources, for which only a fraction of bursts shows oscillations, that too usually for portions of a burst. For ANMXPs, burst oscillation amplitudes are usually close to the accretion-powered pulsation amplitudes \cite{Chakrabartyetal2003,Wattsetal2005,Watts2012}. This is not surprising if both sets of pulsations originate from essentially the same hot spot, perhaps at a magnetic polar cap, as mentioned above. However, burst oscillation amplitudes were found to be somewhat lower than accretion-powered amplitudes for some bursts from XTE J1814--338, which may indicate a larger burning region, perhaps due to a higher degree of fuel spread \cite{Wattsetal2005}. Besides, the decreasing fractional amplitude during the burst rise for SAX J1808.4--3658 implies the thermonuclear flame-spreading, as mentioned above (\cite{BhattacharyyaStrohmayer2006}; see Fig.~\ref{1808_bo}).

Unlike non-AMXPs (except 4U 1636--536) and intermittent AMXPs (see Sect.~\ref{Harmonics}), the harmonic content of burst oscillations from XTE J1814--338 and IGR J17511--3057 has been detected \cite{Strohmayeretal2003,Altamiranoetal2010a}. This supports the hot spot origin of burst oscillations from ANMXPs. However, a difference between the harmonic contents of burst oscillations and accretion-powered pulsations for a given ANMXP may indicate that beaming and other effects could also contribute to this feature \cite{Wattsetal2005}.

While the burst oscillation fractional amplitude for non-AMXPs and intermittent AMXPs increases with photon energy (see Sect.~\ref{Energydependence}), this behaviour appears to be opposite for both burst oscillations and accretion-powered pulsations from ANMXPs \cite{WattsStrohmayer2006,Watts2012}. While this further supports related mechanisms for the two sets of pulsations from ANMXPs, the decrease of the fractional amplitude with energy is not yet fully understood (but see \cite{WattsStrohmayer2006} and references therein).

\subsection{Superburst oscillations}\label{superburst}

Brightness oscillations during an interval of $\sim 800$ s around the peak of a superburst from the known NMXP 4U 1636--536 were observed with {\it RXTE} \cite{StrohmayerMarkwardt2002}. The oscillation frequency was measured to be $\approx 582$~Hz, which is a fraction of a Hz greater than the asymptotic frequency measured from type-I X-ray bursts of this source \cite{Gilesetal2002}. This is similar to intermittent AMXPs, for which the accretion-powered pulsation frequency is $\sim 0.5-1$~Hz greater than the burst oscillation asymptotic frequency (see Sect.~\ref{Frequency}). The superburst oscillation frequency did not drift if the stellar orbital motion is taken into account, and the coherence $Q$ ($> 4.5\times10^5$) was much higher than that observed for burst oscillations. In fact, both these properties are reminiscent of accretion-powered pulsations. The fractional rms amplitude of the superburst oscillations was measured to be $\sim 0.01$, which is lower than those for oscillations found from type-I X-ray bursts from 4U 1636--536 \cite{StrohmayerMarkwardt2002}. Note that if these oscillations were generated only from the non-burst emission, this amplitude would be a few percent, which is typical for AMXPs \cite{PatrunoWatts2012}. A harmonic content of the superburst oscillations was not detected, with a first overtone upper limit of $\sim 6$\% of the fundamental amplitude \cite{StrohmayerMarkwardt2002}. Note that harmonic content has also not been detected for oscillations in individual bursts from non-AMXPs and intermittent AMXPs (see Sect.~\ref{Harmonics}). Even for accretion-powered pulsations, pulse profiles are sinusoidal in most cases, and when there is a harmonic content,
overtones do not usually contribute to the pulsation amplitude by more than $\sim 5$\% \cite{PatrunoWatts2012}. Finally, soft lags were reported for the superburst oscillations \cite{Artigueetal2013}. Note that, while burst oscillations have soft lags in some cases (see Sect.~\ref{Energydependence}), accretion-powered pulsations typically have soft lags \cite{PatrunoWatts2012}.

Therefore, superburst oscillations from 4U 1636--536 were clearly somewhat different from type-I X-ray burst oscillations, particularly if we consider frequency, coherence and fractional amplitude. What could have caused such oscillations? A model based on surface modes (see Sect.~\ref{modes}) could not explain some of the main observed properties \cite{Chambersetal2018}. However, the properties of these oscillations, as discussed above, appear to be consistent with those of accretion-powered pulsations from intermittent AMXPs. Could then 4U 1636--536 be an intermittent AMXP, and were the superburst oscillations accretion-powered pulsations? This could be possible, if the bursting activities around the superburst peak due to $^{12}$C fusion in the deep carbon-rich layer of the ocean (see Sect.~\ref{bursts}) would temporarily uncover a larger magnetic field, which might have been buried by the accreted material. However, given that only one instance of superburst oscillations is known to date, this is a speculation, albeit not unreasonable, at this time.

\section{Burst rise oscillations and thermonuclear flame-spreading}\label{flame-spreading}

Burst oscillations are mostly detected, and their properties, as well as time evolution of these properties, have more extensively been studied for burst decay periods \cite{Watts2012,Galloway2020}. This is because such a period lasts for $\sim 10$~s to tens of seconds, and oscillations are also typically seen at least for a few seconds. On the other hand, burst rise duration is only sub-second to a few seconds, during which burst intensity, spectral properties and oscillation properties evolve rapidly. These make the detection and study of burst rise oscillations more challenging. But while more effort has been given to probe burst decay oscillations, burst rise oscillations are perhaps more important due to the following reasons. (1) The origin of the azimuthally asymmetric brightness pattern is known for burst rise. Thermonuclear flame-spreading gives rise to this pattern, at least for non-AMXPS, and also for some AMXPs, as we will discuss in this section. But we do not have the most basic understanding, i.e., how such a pattern originates, for burst decay oscillations (see Sect.~\ref{Theory}). (2) Whatever is the reason of the asymmetric pattern for burst decay, it should depend on burst ignition and flame-spreading during burst rise (see Sect.~\ref{Theory}). Therefore, it is perhaps not possible to fully understand burst decay oscillations without understanding burst rise oscillations. (3) Because of the above points, and an expanding hot spot origin, burst rise oscillations are likely a more reliable tool to constrain neutron star parameters, and hence to probe the supra-nuclear density degenerate core matter of neutron stars (\cite{Bhattacharyya2010}; see Sect.~\ref{Introduction}).

\subsection{Theory of flame-spreading}\label{flame_theory}

A thermonuclear burst should be ignited at some point on the neutron star surface, because, given that burst durations are much shorter than time separations between bursts, it is very unlikely that ignition conditions can be met simultaneously all over the surface \cite{Shara1982,CummingBildsten2000}. It was noted that bursts should preferentially be ignited at or very close to the equator, because the spin-induced reduced gravity should lead to a larger accumulation of the accreted matter at the equator \cite{Spitkovskyetal2002}. However, for higher accretion rates, ignition may occur at higher latitudes \cite{CooperNarayan2007}.

After ignition, a thermonuclear flame-spreading on the stellar surface to burn the fuel, i.e., the accumulated accreted matter, is expected. This spreading could happen via deflagration \cite{FryxellWoosley1982b} or detonation \cite{FryxellWoosley1982a,Zingaleetal2001}. Detonation could be possible, if the nuclear burning timescale is less than the vertical sound crossing time, which is not expected for NMXPs \cite{Spitkovskyetal2002}. Therefore, flames should spread on the stellar surface by deflagration.

A theoretical study of flame-spreading on NMXPs, considering the effects of the Coriolis force and the puffed up burning layers, were reported in 2002 \cite{Spitkovskyetal2002}. According to this study, initially flames should spread by geostrophic flow, rapidly and almost isotropically, with a typical speed $\vartheta_{\rm geostrophic} \approx \sqrt{gh} \sim 4500$~km s$^{-1}$. Here, $g$ and $h$ are the gravitational acceleration and the burning region scale height, respectively. However, when the burning region becomes relatively large, with the
Rossby number less than 1, the Coriolis force becomes important, and flames start spreading with a significantly slower ageostrophic speed. This speed $\vartheta_{\rm flame}$ is $\sim \sqrt{gh}/ft_{\rm n} \sim 5-10$~km s$^{-1}$~km s$^{-1}$ for a weak turbulent viscosity, and $\sim \sqrt{gh/ft_{\rm n}} \sim 100-300$~km s$^{-1}$~km s$^{-1}$ for a dynamically important turbulent viscosity \cite{Spitkovskyetal2002,ChakrabortyBhattacharyya2014}. Here, $t_{\rm n}$ is the nuclear burning time scale and $f = 2 \Omega \sin{\theta}$ is the Coriolis parameter, with $\Omega$ being the neutron star angular speed and $\theta$ being the latitude of the burning front. Therefore, the flame speed is latitude dependent and anisotropic \cite{Spitkovskyetal2002}. A few further theoretical studies on flame propagation have been reported since 2002 (e.g., \cite{Cavecchietal2013,Cavecchietal2015,Cavecchietal2016,CavecchiSpitkovsky2019,Eidenetal2020}), some of which have considered the effects of magnetic field and instability, and used simulations. Note that the expanding burning region, which remains somewhat confined by the  Coriolis force, should naturally give rise to brightness oscillations observed during the burst rise. The above-mentioned theoretical results are important to understand various aspects of such burst rise oscillations, as we will discuss in Sect.~\ref{Further}.

\subsection{Evidence of flame-spreading }\label{Further}

An expanding burning region on a spinning neutron star should give rise to brightness oscillations with a decreasing fractional amplitude, as an increasingly larger fraction of such a region is expected to remain visible throughout the stellar spin period. Such an amplitude evolution of burst rise oscillations should, therefore, be a signature of thermonuclear flame-spreading. In addition, since the burst X-ray spectrum is typically well described with a blackbody, the burst emission area should be proportional to the best-fit blackbody normalization value. An increase of this normalization with time during burst rise could, therefore, provide an independent evidence of flame-spreading, although systematics might affect the inferred burst emission area \cite{Bhattacharyyaetal2010}. Hence, the study of evolution of burst intensity, spectral properties and oscillation properties during the burst rise can be tools to probe flame-spreading. However, in order to use these observational aspects as tools, one needs to divide the burst rise period into a number of smaller time bins, and significantly detect features like oscillations and measure their properties in each bin. This is an extremely  challenging goal, given a typical burst rise duration is only $\sim 1$~s.

In 1997, an evidence of decreasing fractional amplitude of burst rise oscillations was reported for the NMXP 4U 1728--34 \cite{Strohmayeretal1997a}. Soon after this, more indications of plausible flame-spreading were found from oscillations and spectral analyses during burst rise (e.g., \cite{Strohmayeretal1998a,vanStraatenetal2001}). In 2005, a better than $3\sigma$ detection of harmonic content of oscillations during the initial part of burst rise was reported for the NMXP 4U 1636--536 \cite{BhattacharyyaStrohmayer2005}. This and the finding of a gradual decrease of the harmonic power is expected for an expanding burning region. During 2006--07, observational indications of flame-spreading were reported for multiple bursts from two NMXPs. One of them was the ANMXP SAX J1808.4--3658, for which the oscillation frequency was found to increase rapidly by a few Hz and overshoot the pulsar frequency (see Sect.~\ref{AMXP}). With a detailed analysis it appeared that during the rise of one burst, the oscillation frequency could have increased by a few Hz in just first $\sim 0.1$ s, then decreased, and then increased again by a few Hz, all in just $\sim 0.5$~s \cite{BhattacharyyaStrohmayer2006}. For another burst from the same source, the oscillation frequency increased by a few Hz and appeared to overshoot the pulsar frequency in the first $\sim 0.2$ s, and then decreased and remained almost same for the next $\sim 0.5$~s \cite{BhattacharyyaStrohmayer2006}. During the first $\sim 0.2$ s, when this frequency rapidly increased, the oscillation fractional amplitude quickly decreased, and the apparent radius, which is proportional to the square-root of the burst emission area, increased fast (see Fig.~\ref{1808_bo}). For the next $\sim 0.5$~s, both the amplitude and the apparent radius maintained near-constant values. Such an evolution and the anti-correlation between the oscillation fractional amplitude and the burst emission area are expected for the flame-spreading during burst rise. For another burst from the same source, which was a precursor to a big burst, a joint spectral and temporal modelling indicated an expanding burning region \cite{BhattacharyyaStrohmayer2007a}. Analyses of burst oscillations and X-ray spectra also indicated thermonuclear flame-spreading, and a temporary burning front stalling, for non-PRE double-peaked bursts from 4U 1636--536 \cite{BhattacharyyaStrohmayer2006a,BhattacharyyaStrohmayer2006b}.

\begin{figure}[t]
\sidecaption
\includegraphics[scale=.42]{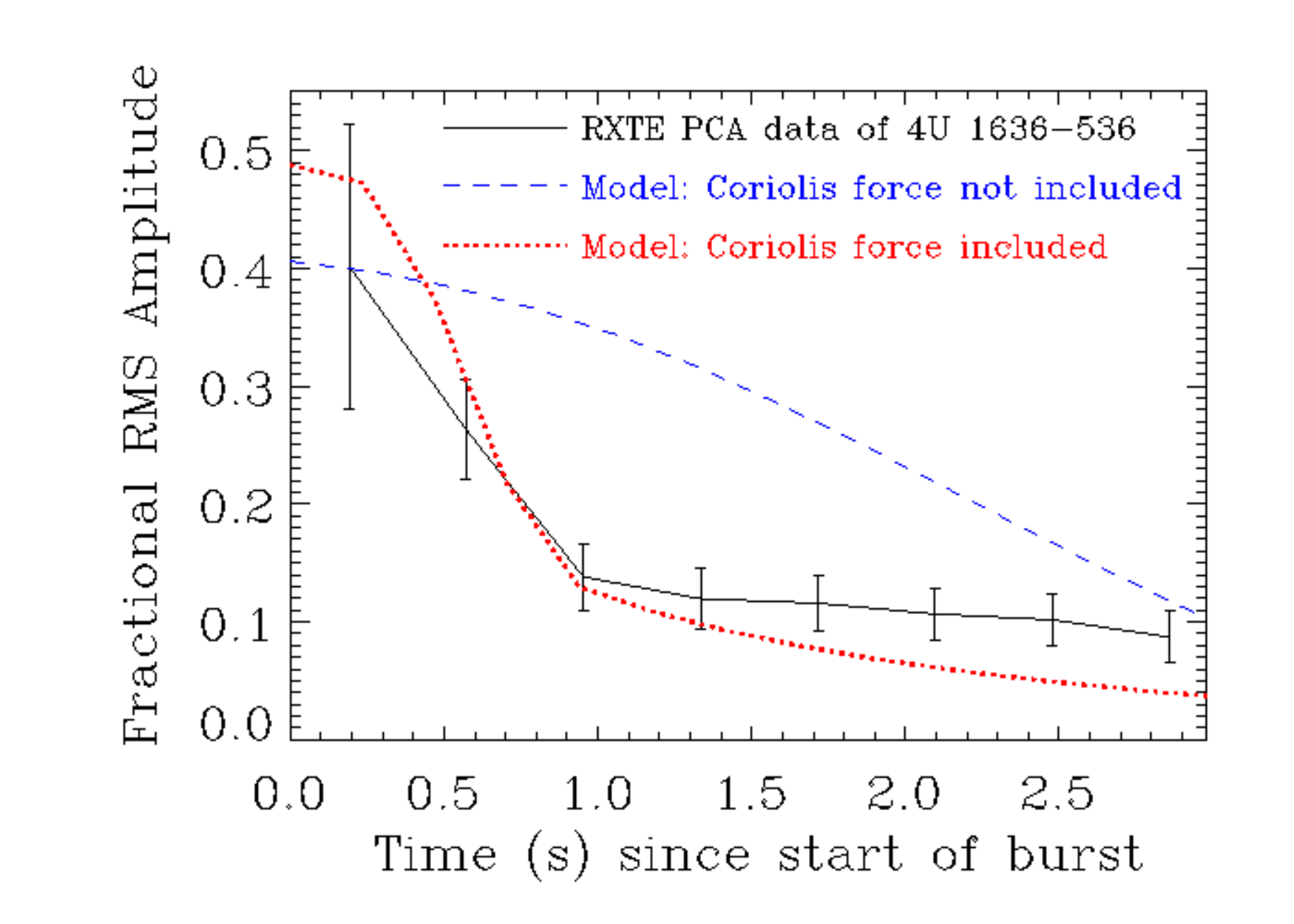}
\caption{Evolution of fractional rms amplitude of burst oscillations (black solid line with error bars) during the rise of a thermonuclear X-ray burst observed from the neutron star LMXB 4U 1636--536 with the {\it Rossi X-ray Timing Explorer} ({\it RXTE}) satellite \cite{BhattacharyyaStrohmayer2007}. Two example models (not fit to data) -- one for a uniformly expanding circular hot spot without considering the Coriolis force effect (blue dashed curve) and another for flame spreading including the salient features of the Coriolis force (red dotted line) -- are shown (see Sect.~\ref{Further}).}
\label{flame}
\end{figure}

It was found that, if the burning region expands isotropically, the oscillation fractional amplitude first decreases slowly, and then more rapidly, making the evolution profile upward convex \cite{BhattacharyyaStrohmayer2007}. But if this region expands in a latitude-dependent and anisotropic manner, which captures some features of the effects of the Coriolis force (see Sect.~\ref{flame_theory}), the oscillation fractional amplitude first decreases rapidly, and then rather slowly, making the evolution profile upward concave \cite{BhattacharyyaStrohmayer2007}. It was reported for two bursts from 4U 1636--536 and SAX J1808.4--3658 that a model of isotropic spreading cannot explain the observed oscillation fractional amplitude evolution, while a model including the salient features of the Coriolis force can, at least qualitatively \cite{BhattacharyyaStrohmayer2007}. This is shown in Fig.~\ref{flame}.

All the indications of thermonuclear flame-spreading mentioned above are for a few individual bursts, and a quantitative, systematic and extensive study to establish a plausible decreasing trend of oscillation fractional amplitude during burst rise was missing. Besides, when an observed fractional amplitude evolution during burst rise was studied, a sufficiently significant detection of oscillations in each independent time bin was not always considered. In 2014, a comprehensive study using the rising portions of the {\it RXTE}-observed  thermonuclear X-ray bursts from 10 NMXPs was reported \cite{ChakrabortyBhattacharyya2014}. The rise times of bursts from all 10 sources were divided into $0.25$ s independent bins. Such a short bin was ideal to track oscillation fractional amplitude evolution, particularly at the beginning of burst rise, when the amplitude appears to evolve quickly. For a given time bin, the fractional rms amplitude was measured and an error was estimated, only if oscillations in that bin was detected with at least $3\sigma$ significance. Otherwise, an upper limit of the fractional rms amplitude was estimated. These made the analyses systematic and uniform across the bursts and sources. Moreover, the stringent detection criterion, given that a short 0.25 s bin has a small number of total counts, contributed
to the reliability of the conclusions. With this criterion, oscillations in at least one bin was detected during the rise of 51 bursts. By fitting the fractional rms amplitude evolution, including upper limits, with empirical models, the decreasing trend was detected with $4\sigma$ significance for the best case, which is by far the strongest evidence of thermonuclear flame-spreading by this method till now \cite{ChakrabortyBhattacharyya2014}. Moreover, an opposite trend was not found in any burst. A quantitative analysis also revealed that the fractional amplitude profiles are upward concave, which implied latitude-dependent flame speeds, likely because the effects of the Coriolis force are important \cite{ChakrabortyBhattacharyya2014}. This is what was theoretically predicted (see Sect.~\ref{flame_theory}). Besides, even without considering the details, the finding of flame-spreading during $\sim 2$~s implied an average flame speed of $\sim 15$ km s$^{-1}$, which indicated a weak turbulent viscosity (see Sect.~\ref{flame_theory}).

It is, however, puzzling, why not all bursts, not even all bursts from NMXPs, show oscillations, at least during the burst rise. This could be somewhat understood based on the ignition location \cite{ChakrabortyBhattacharyya2014,MaurerWatts2008}. As mentioned in Sect.~\ref{flame_theory}, most bursts are likely ignited at or very close to the equator. Therefore, a very large equatorial flame speed (see Sect.~\ref{flame_theory}) should ignite the entire equatorial belt in a tiny fraction of a second, making the burning region azimuthally symmetric. The subsequent symmetric northward and southward flame-spreading naturally should not give rise to oscillations in most bursts. However, if ignition happens at higher latitudes, possibly for higher accertion rates \cite{CooperNarayan2007}, an azimuthally asymmetric burning region may survive for an observable duration, and burst oscillations can be detected. This is perhaps why burst oscillations are preferentially seen at higher accretion rates (see Sect.~\ref{Amplitude}). How long burst rise oscillations can be observed would then depend on exactly at what latitude and in which hemisphere the burst has been ignited. But, if a burst is ignited at a very high latitude, oscillation amplitude could be very small depending on the observer's line of sight, and the burning region may become azimuthally almost symmetric soon after the ignition \cite{BhattacharyyaStrohmayer2006b}. Note that for ANMXPs, it is likely that a burst is ignited at a magnetic polar cap because of its higher temperature, and then the flames may spread for some sources, like SAX J1808.4--3658 \cite{BhattacharyyaStrohmayer2006}, and may not spread much for some other sources, like XTE J1814--338 \cite{Strohmayeretal2003,Wattsetal2005}.

\section{What causes burst decay oscillations?}\label{Theory}

Oscillations during burst decay of ANMXPs could be caused by an azimuthal brightness asymmetry due to a magnetic polar cap (see Sect.~\ref{AMXP}), although this is not yet established. It is, however, more challenging to explain the observed burst decay oscillations for non-AMXPs and intermittent AMXPs. This is because, after a likely thermonuclear flame-spreading all over the neutron star surface during burst rise, it is not clear how an azimuthally asymmetric brightness pattern survives on the stellar surface. Therefore, while several models, including those based on surface oscillation modes \cite{Heyl2004}, cooling wake \cite{CummingBildsten2000,Mahmoodifaretal2016}, vortices \cite{Spitkovskyetal2002} and convective patterns \cite{Garciaetal2019}, have been proposed, the cause of the asymmetry is not yet known with certainty. Here we briefly describe two main models, based on surface modes and cooling wake.

\subsection{Surface modes}\label{modes}

Thermonuclear flame-spreading during the rise of a burst could excite waves in the ocean and the upper atmosphere of the neutron star, which could persist throughout a large fraction of the burst decay period \cite{Heyl2004}. Height differences due to these waves could give rise to an azimuthally asymmetric temperature, and hence brightness, pattern, and therefore, burst decay oscillations \cite{Watts2012}. As the stellar surface cools down, the frequency of the waves may evolve, which could, in principle, explain the observed oscillation frequency drift (see Sect.~\ref{Frequency}). Different families of modes could be excited. But it was found that observations are best explained for the buoyant Rossby modes or r-modes, which are driven by a latitudinal variation of the Coriolis force, and the buoyancy \cite{Heyl2004,Watts2012}. It is, however, not clear, why, among all the plausible modes, only some specific r-modes would be excited. Moreover, while the surface mode model can explain the observed absence of a significant harmonic content and the energy-dependence of amplitude \cite{Heyl2004,LeeStrohmayer2005}, this model cannot entirely explain the magnitudes of frequency drift and fractional amplitude. The predicted frequency drift is typically much higher than that observed \cite{Watts2012,Mahmoodifaretal2016}, although inclusion of the relativistic effects and use of improved models for the ocean cooling can reduce this discrepancy \cite{ChambersWatts2020}. Besides, while this surface mode model can explain the lower values of observed burst decay oscillation fractional amplitudes, it is not clear if this model can explain higher observed amplitude values \cite{Mahmoodifaretal2016}. Particularly, it is not likely that the surface mode model would be able to explain fractional rms amplitudes as high as $\sim 0.4-0.5$, observed from 4U 1728--34 (\cite{Mahmoodifaretal2019}; see Sect.~\ref{Amplitude}).

\subsection{Cooling wake}\label{wake}

A `cooling wake', i.e., a temperature asymmetry during the cooling of the stellar surface, is a natural consequence of thermonuclear flame-spreading during the burst rise. Such a cooling wake could give rise to an azimuthally asymmetric brightness pattern, which might explain burst decay oscillations \cite{CummingBildsten2000}. However, it has been found that for a `canonical' cooling, for which either each portion on the stellar surface heats up and cools down in the same way or the cooling timescale varies with latitude, unusually long burst rise times are required to produce the higher values of observed oscillation fractional amplitudes \cite{Mahmoodifaretal2016}. A canonical cooling also predicts a correlation between burst rise times and the decay oscillation amplitudes \cite{Mahmoodifaretal2016}, which is not observed \cite{Chakrabortyetal2017}. On the other hand, for an `asymmetric' cooling model, for which different parts of the stellar surface cool at considerably different rates, higher oscillation fractional amplitudes can be produced, and there is no requirement of a correlation between burst rise times and the decay oscillation amplitudes \cite{Mahmoodifaretal2016}. While the asymmetric cooling wake mechanism has so far not been studied much, it is strongly connected to the thermonuclear flame-spreading during burst rise, and together they might provide a natural and self-consistent mechanism for both rise and decay oscillations \cite{Mahmoodifaretal2016}.  However, the origin of an asymmetry in the cooling is not yet understood. 

\section{Conclusion}\label{Conclusion}

Observations, primarily with the {\it RXTE} satellite, have confirmed the following points about burst oscillations: (1) the oscillations originate due to an azimuthally asymmetric brightness pattern on the surface of a spinning neutron star \cite{StrohmayerBildsten2006,Watts2012}, (2) the oscillation frequencies are very close to (typically within $1$\% of) the stellar spin frequency \cite{Chakrabartyetal2003}, and (3) oscillations during burst rise are mainly caused by thermonuclear flame-spreading on the stellar surface, except for some AMXPs \cite{Strohmayeretal1997a,ChakrabortyBhattacharyya2014}. Nevertheless, many aspects of burst oscillations are still not well-understood. The main problem is to understand the origin of oscillations during burst decay, when the fuel on the entire neutron star surface is expected to be already burnt. But, even for burst rise, several aspects, such as what causes the frequency drift (see Figs.~\ref{1636_bo} and \ref{1808_bo}) and why hard lags are observed \cite{Chakrabortyetal2017}, are not understood. Some puzzling aspects of burst oscillations are why they are not detected from all bursting neutron star LMXBs, why they are not detected from all bursts for most NMXPs and why they are intermittent and short-lived for many bursts. While we have discussed some of these points for burst rise in Sect.~\ref{Further} (see also \cite{ChakrabortyBhattacharyya2014}), in order to understand burst oscillations, first we need to either detect or at least put stringent upper limits on this feature in sub-second time bins throughout bursts, including in the time bins at the very beginning of the burst rise, when burst count rates are relatively small. Then it is required to measure all burst properties, such as, oscillation frequency, amplitudes of the fundamental and plausible overtones, energy-dependence of amplitudes and phase, etc., in conjunction with burst spectral properties, in all such time bins, which would give the time evolution of these properties and likely correlations among them. These will require observations with future large area X-ray instruments with a sufficient time resolution. Note that the accretion-powered emission may evolve during a burst, and hence could introduce significant systematics into measured burst properties \cite{Worpeletal2013,Bhattacharyyaetal2018,ChakrabortyBhattacharyya2014}. In order to reliably measure these properties, it is therefore essential to characterize the accretion-powered emission, and hence the broadband spectra of NMXPs. These could perhaps be achieved by simultaneous observations with two mutually well-calibrated instruments, one for softer X-rays, like a larger area version of {\it NICER} \cite{GendreauArzoumanian2017}, and another for harder X-rays, like a larger area and better spectral resolution version of {\it AstroSat} Large Area X-ray Proportional Counters (LAXPC; \cite{Antiaetal2017}). Another technique, i.e.,  pulse phase polarimetry, could also be used to probe burst oscillations in the future \cite{Ghoshetal2013}.

Burst oscillation light curves should be affected by various physical effects, such as Doppler shift, special relativistic beaming, gravitational red shift and light bending, etc. Therefore, a modelling of such phase-folded and energy-resolved light curves can be useful to estimate source parameters, including neutron star mass and radius, and hence to understand the supra-nuclear density degenerate stellar core matter \cite{MillerLamb1998,Nathetal2002,Bhattacharyyaetal2005,Loetal2013}. In addition, studies of burst oscillations can be very useful to probe the extreme environments of the atmosphere, and even the crust, of neutron stars, which are affected by strong gravity, high magnetic field, intense radiation and rapid stellar spin. As an example, an observational evidence from oscillations that flame-spreading happens throughout the burst rise of $\sim 2$~s could put a constraint on the viscosity (see Sect.~\ref{Further}). However note that, only when the origin and physics of burst oscillations are sufficiently well-understood, NMXPs can be reliably used to probe the fundamental physics of neutron star interior and surface.



\bibliographystyle{spphys.bst}
\bibliography{nmxpbiblio}

\begin{thebibliography}{100}
\providecommand{\url}[1]{{#1}}
\providecommand{\urlprefix}{URL }
\expandafter\ifx\csname urlstyle\endcsname\relax
  \providecommand{\doi}[1]{DOI \discretionary{}{}{}#1}\else
  \providecommand{\doi}{DOI \discretionary{}{}{}\begingroup
  \urlstyle{rm}\Url}\fi

\bibitem{Bhattacharyaetal1991}
D.~{Bhattacharya}, E.P.J. {van den Heuvel}, {PhR} \textbf{203}(1-2), 1 (1991).
\newblock \doi{10.1016/0370-1573(91)90064-S}

\bibitem{PatrunoWatts2012}
A.~Patruno, A.L. Watts, in \emph{{Timing neutron stars: pulsations,
  oscillations and explosions}}, ed. by T.~Belloni, M.~Mendez, C.M. Zhang
  (ASSL, Springer, 2012)

\bibitem{Watts2012}
A.L. {Watts}, {ARA\&A} \textbf{50}, 609 (2012).
\newblock \doi{10.1146/annurev-astro-040312-132617}

\bibitem{Bhattacharyya2010}
S.~Bhattacharyya, Advances in Space Research \textbf{45}, 949 (2010).
\newblock \doi{10.1016/j.asr.2010.01.010}

\bibitem{StrohmayerBildsten2006}
T.E. Strohmayer, L.~Bildsten, in \emph{{Compact Stellar X-ray Sources}},
  vol.~39, ed. by W.H.G. Lewin, M.~van~der Klis (Cambridge University Press,
  Cambridge, 2006), pp. 113--156

\bibitem{Galloway2008}
D.K. Galloway, M.P. Muno, J.M. Hartman, D.~Psaltis, D.~Chakrabarty, ApJS
  \textbf{179}, 360 (2008).
\newblock \doi{10.1086/592044}

\bibitem{Galloway2020}
D.K. Galloway, J.~in't Zand, J.~Chenevez, H.~W{\"o}rpel, L.~Keek, L.~Ootes,
  A.L. Watts, L.~Gisler, C.~Sanchez-Fernandez, E.~Kuulkers, ApJS \textbf{249},
  32 (2020).
\newblock \doi{10.3847/1538-4365/ab9f2e}

\bibitem{Grindlay1976}
J.~Grindlay, H.~Gursky, H.~Schnopper, D.R. Parsignault, J.~Heise, A.C.
  Brinkman, J.~Schrijver, ApJ \textbf{205}, L127 (1976).
\newblock \doi{10.1086/182105}

\bibitem{Belian1976}
R.D. Belian, J.P. Conner, W.D. Evans, ApJ \textbf{206}, L135 (1976).
\newblock \doi{10.1086/182151}

\bibitem{Swank1977}
J.H. Swank, R.H. Becker, E.A. Boldt, S.S. Holt, S.H. Pravdo, P.J. Serlemitsos,
  ApJ \textbf{212}, L73 (1977).
\newblock \doi{10.1086/182378}

\bibitem{Hoffmanetal1977a}
J.A. Hoffman, W.H.G. Lewin, J.~Doty, ApJ \textbf{217}, L23 (1977).
\newblock \doi{10.1086/182531}

\bibitem{Joss1977}
P.C. Joss, Nature \textbf{270}, 310 (1977).
\newblock \doi{10.1038/270310a0}

\bibitem{LambLamb1978}
D.Q. Lamb, F.K. Lamb, ApJ \textbf{220}, 291 (1978).
\newblock \doi{10.1086/155905}

\bibitem{Joss1979}
P.C. Joss, Comm. Astrophys. \textbf{8}, 109 (1979)

\bibitem{zand2008}
J.J.M. in't Zand, C.G. Bassa, P.G. Jonker, L.~Keek, F.~Verbunt, M.~M\'endez,
  C.B. Markwardt, A\&A \textbf{485}, 183 (2008).
\newblock \doi{10.1051/0004-6361:200809361}

\bibitem{Bildsten2000}
L.~Bildsten, in \emph{{AIP Conference Proceedings}}, vol. 522 (2000), vol. 522,
  pp. 359--369.
\newblock \doi{10.1063/1.1291736}

\bibitem{Fowler1965}
W.A. Fowler, F.~Hoyle, \emph{{Nucleosynthesis in massive stars and supernovae}}
  (University of Chicago Press, Chicago, 1965)

\bibitem{Schatz2001}
H.~Schatz, A.~Aprahamian, V.~Barnard, L.~Bildsten, A.~Cumming, M.~Ouellette,
  T.~Rauscher, F.K. Thielemann, M.~Wiescher, PRL \textbf{86}, 3471 (2001).
\newblock \doi{10.1103/PhysRevLett.86.3471}

\bibitem{Ayasli1982}
S.~Ayasli, P.C. Joss, ApJ \textbf{256}, 637 (1982).
\newblock \doi{10.1086/159940}

\bibitem{Taam1996}
R.E. Taam, S.E. Woosley, D.Q. Lamb, ApJ \textbf{459}, 271 (1996).
\newblock \doi{10.1086/176890}

\bibitem{Gallowayetal2004}
D.K. Galloway, A.~Cumming, E.~Kuulkers, L.~Bildsten, D.~Chakrabarty, R.E.
  Rothschild, ApJ \textbf{601}, 466 (2004).
\newblock \doi{10.1086/380445}

\bibitem{ChakrabortyBhattacharyya2011}
M.~Chakraborty, S.~Bhattacharyya, ApJ \textbf{730}, L23 (2011).
\newblock \doi{10.1088/2041-8205/730/2/L23}

\bibitem{Chakrabortyetal2011}
M.~Chakraborty, S.~Bhattacharyya, A.~Mukherjee, MNRAS \textbf{418}, 490 (2011).
\newblock \doi{10.1111/j.1365-2966.2011.19499.x}

\bibitem{Linaresetal2012}
M.~Linares, D.~Altamirano, D.~Chakrabarty, A.~Cumming, L.~Keek, ApJ
  \textbf{748}, 82 (2012).
\newblock \doi{10.1088/0004-637X/748/2/82}

\bibitem{Spitkovskyetal2002}
A.~Spitkovsky, Y.~Levin, G.~Ushomirsky, ApJ \textbf{566}, 1018 (2002).
\newblock \doi{10.1086/338040}

\bibitem{Woosleyetal2004}
S.E. Woosley, A.~Heger, A.~Cumming, R.D. Hoffman, J.~Pruet, T.~Rauscher, J.L.
  Fisker, H.~Schatz, B.A.B.M. Wiescher, ApJS \textbf{151}, 75 (2004).
\newblock \doi{10.1086/381533}

\bibitem{Weinbergetal2006}
N.N. Weinberg, L.~Bildsten, H.~Schatz, ApJ \textbf{639}, 1018 (2006).
\newblock \doi{10.1086/499426}

\bibitem{Maloneetal2011}
C.M. Malone, A.~Nonaka, A.S. Almgren, J.B. Bell, M.~Zingale, ApJ \textbf{728},
  118 (2011).
\newblock \doi{10.1088/0004-637X/728/2/118}

\bibitem{Garciaetal2018}
F.~Garcia, F.R.N. Chambers, A.L. Watts, PhRvF \textbf{3}, 123501 (2018).
\newblock \doi{10.1103/PhysRevFluids.3.123501}

\bibitem{KeekHeger2017}
L.~Keek, A.~Heger, ApJ \textbf{842}, 113 (2017).
\newblock \doi{10.3847/1538-4357/aa7748}

\bibitem{Kapteinetal2000}
R.G. Kaptein, J.J.M. in't Zand, E.~Kuulkers, F.~Verbunt, J.~Heise,
  R.~Cornelisse, A\&A \textbf{358}, L71 (2000)

\bibitem{intZandetal2002}
J.J.M. in't Zand, F.~Verbunt, E.~Kuulkers, C.B. Markwardt, A.~Bazzano,
  M.~Cocchi, R.~Cornelisse, J.~Heise, L.~Natalucci, P.~Ubertini, A\&A
  \textbf{389}, L43 (2002).
\newblock \doi{10.1051/0004-6361:20020631}

\bibitem{intZandetal2005}
J.J.M. in't Zand, A.~Cumming, M.~van der Sluys~andF. Verbunt, O.R. Pols, A\&A
  \textbf{441}, 675 (2005).
\newblock \doi{10.1051/0004-6361:20053002}

\bibitem{Cornelisse2000}
R.~Cornelisse, J.~Heise, E.~Kuulkers, F.~Verbunt, J.J.M. in't Zand, A\&A
  \textbf{357}, L21 (2000)

\bibitem{Kuulkers2004}
E.~Kuulkers, NuPhS \textbf{132}, 466 (2004).
\newblock \doi{10.1016/j.nuclphysbps.2004.04.081}

\bibitem{intZandetal2004}
J.J.M. in't Zand, R.~Cornelisse, A.~Cumming, A\&A \textbf{426}, 257 (2004).
\newblock \doi{10.1051/0004-6361:20040522}

\bibitem{Kuulkers2005}
E.~Kuulkers, ATel \textbf{483}, 1 (2005)

\bibitem{Keek2008}
L.~Keek, J.J.M. in't Zand, E.~Kuulkers, A.~Cumming, E.F. Brown, M.~Suzuki, A\&A
  \textbf{479}, 177 (2008).
\newblock \doi{10.1051/0004-6361:20078464}

\bibitem{WoosleyTaam1976}
S.E. Woosley, R.E. Taam, Nature \textbf{263}, 101 (1976).
\newblock \doi{10.1038/263101a0}

\bibitem{TaamPicklum1978}
R.E. Taam, R.E. Picklum, ApJ \textbf{224}, 210 (1978).
\newblock \doi{10.1086/156367}

\bibitem{BrownBildsten1998}
E.F. Brown, L.~Bildsten, ApJ \textbf{496}, 915 (1998).
\newblock \doi{10.1086/305419}

\bibitem{bac82}
D.C. {Backer}, S.R. {Kulkarni}, C.~{Heiles}, M.M. {Davis}, W.M. {Goss}, Nature
  \textbf{300}, 615 (1982).
\newblock \doi{10.1038/300615a0}

\bibitem{rad82}
V.~{Radhakrishnan}, G.~{Srinivasan}, Current Science \textbf{51}, 1096 (1982)

\bibitem{alp82}
M.A. {Alpar}, A.F. {Cheng}, M.A. {Ruderman}, J.~{Shaham}, Nature \textbf{300},
  728 (1982).
\newblock \doi{10.1038/300728a0}

\bibitem{Jahodaetal2006}
K.~Jahoda, C.B. Markwardt, Y.~Radeva, A.H. Rots, M.J. Stark, J.H. Swank, T.E.
  Strohmayer, W.~Zhang, ApJS \textbf{163}, 401 (2006).
\newblock \doi{10.1086/500659}

\bibitem{Strohmayeretal1996}
T.E. Strohmayer, W.~Zhang, J.H. Swank, A.~Smale, L.~Titarchuk, C.~Day, U.~Lee,
  ApJ \textbf{469}, L9 (1996).
\newblock \doi{10.1086/310261}

\bibitem{Strohmayeretal1999}
T.E. Strohmayer, J.H. Swank, W.~Zhang, NuPhS \textbf{69}, 129 (1999).
\newblock \doi{10.1016/S0920-5632(98)00195-9}

\bibitem{Strohmayeretal2001}
T.E. Strohmayer, Advances in Space Research \textbf{28}, 511 (2001).
\newblock \doi{10.1016/S0273-1177(01)00422-7}

\bibitem{Strohmayeretal1998}
T.E. Strohmayer, W.~Zhang, J.H. Swank, I.~Lapidus, ApJ \textbf{503}, L147
  (1998).
\newblock \doi{10.1086/311545}

\bibitem{vanderKlis2006}
M.~van~der Klis, in \emph{{Compact Stellar X-ray Sources}}, vol.~39, ed. by
  W.H.G. Lewin, M.~van~der Klis (Cambridge University Press, Cambridge, 2006),
  pp. 39--112

\bibitem{Miller1999}
M.C. Miller, ApJ \textbf{515}, L77 (1999).
\newblock \doi{10.1086/311970}

\bibitem{Chakrabartyetal2003}
D.~Chakrabarty, E.H. Morgan, M.P. Muno, D.K. Galloway, R.~Wijnands, M.~van~der
  Klis, C.B. Markwardt, Nature \textbf{424}, 42 (2003).
\newblock \doi{10.1038/nature01732}

\bibitem{Bhattacharyyaetal2006}
S.~Bhattacharyya, T.E. Strohmayer, C.B. Markwardt, J.H. Swank, ApJ
  \textbf{639}, L31 (2006).
\newblock \doi{10.1086/501438}

\bibitem{Bhattacharyyaetal2005}
S.~Bhattacharyya, T.E. Strohmayer, M.C. Miller, C.B. Markwardt, ApJ
  \textbf{619}, 483 (2005).
\newblock \doi{10.1086/426383}

\bibitem{Hartmanetal2003}
J.M. Hartman, D.~Chakrabarty, D.K. Galloway, M.P. Muno, P.~Savov, M.~Mendez,
  S.~van Straaten, T.D. Salvo, Bull. Am. Astron. Soc. \textbf{35}, 865 (2003)

\bibitem{Kaaretetal2002}
P.~Kaaret, J.J.M. in~'t Zand, J.~Heise, J.A. Tomsick, ApJ \textbf{575}, 1018
  (2002).
\newblock \doi{10.1086/341336}

\bibitem{Strohmayeretal1997}
T.E. Strohmayer, K.~Jahoda, A.B. Giles, U.~Lee, ApJ \textbf{486}, 355 (1997).
\newblock \doi{10.1086/304522}

\bibitem{Strohmayeretal1998a}
T.E. Strohmayer, W.~Zhang, J.H. Swank, N.E. White, I.~Lapidus, ApJ
  \textbf{498}, L135 (1998).
\newblock \doi{10.1086/311322}

\bibitem{Gilesetal2002}
A.B. Giles, K.M. Hill, T.E. Strohmayer, N.~Cummings, ApJ \textbf{568}, 279
  (2002).
\newblock \doi{10.1086/338890}

\bibitem{StrohmayerMarkwardt2002}
T.E. Strohmayer, C.B. Markwardt, ApJ \textbf{577}, 337 (2002).
\newblock \doi{10.1086/342152}

\bibitem{Wijnandsetal2001}
R.~Wijnands, T.~Strohmayer, L.M. Franco, ApJ \textbf{549}, L71 (2001).
\newblock \doi{10.1086/319128}

\bibitem{Galloway2010}
D.K. Galloway, J.~Lin, D.~Chakrabarty, .~J.~M.~Hartman3, ApJ \textbf{711}, L148
  (2010).
\newblock \doi{10.1088/2041-8205/711/2/L148}

\bibitem{Zhangetal1998}
W.~Zhang, K.~Jahoda, R.L. Kelley, T.E. Strohmayer, J.H. Swank, S.N. Zhang, ApJ
  \textbf{495}, L9 (1998).
\newblock \doi{10.1086/311210}

\bibitem{Bilousetal2018}
A.V. Bilous, A.L. Watts, D.K. Galloway, J.J.M. in't Zand, ApJ \textbf{862}, L4
  (2018).
\newblock \doi{10.3847/2041-8213/aad09c}

\bibitem{Smithetal1997}
D.A. Smith, E.H. Morgan, H.~Bradt, ApJ \textbf{479}, L137 (1997).
\newblock \doi{10.1086/310604}

\bibitem{Linaresetal2011}
M.~Linares, D.~Altamirano, A.W.T. Strohmayer, D.~Chakrabarty, A.~Patruno,
  M.~van~der Klis, R.~Wijnands, P.~Casella, M.~Armas-Padilla, Y.~Cavecchi,
  N.~Degenaar, M.~Kalamkar, R.~Kaur, Y.~Yang, N.~Rea, The Astronomer's Telegram
  \textbf{3568}, 1 (2011)

\bibitem{ChakrabortyBhattacharyya2012}
M.~Chakraborty, S.~Bhattacharyya, MNRAS \textbf{422}, 2351 (2012).
\newblock \doi{10.1111/j.1365-2966.2012.20786.x}

\bibitem{Wattsetal2009}
A.L. Watts, D.~Altamirano, M.~Linares, A.~Patruno, P.~Casella, Y.~Cavecchi,
  N.~Degenaar, N.~Rea, P.~Soleri, M.~van~der Klis, R.~Wijnands, ApJ
  \textbf{698}, L174 (2009).
\newblock \doi{10.1088/0004-637X/698/2/L174}

\bibitem{BilousWatts2019}
A.V. Bilous, A.L. Watts, ApJS \textbf{245}, 19 (2019).
\newblock \doi{10.3847/1538-4365/ab2fe1}

\bibitem{Markwardtetal1999}
C.B. Markwardt, T.E. Strohmayer, J.H. Swank, ApJ \textbf{512}, L125 (1999).
\newblock \doi{10.1086/311886}

\bibitem{Strohmayeretal2003}
T.E. Strohmayer, C.B. Markwardt, J.H. Swank, J.~in't Zand, ApJ \textbf{596},
  L67 (2003).
\newblock \doi{10.1086/379158}

\bibitem{Altamiranoetal2010}
D.~Altamirano, M.~Linares, A.~Patruno, N.~Degenaar, R.~Wijnands, M.~Klein-Wolt,
  M.~van~der Klis, C.~Markwardt, J.~Swank, MNRAS \textbf{401}, 223 (2010).
\newblock \doi{10.1111/j.1365-2966.2009.15627.x}

\bibitem{Patruno2013}
A.~Patruno, The Astronomer's Telegram \textbf{5068}, 1 (2013)

\bibitem{pap13}
A.~{Papitto}, C.~{Ferrigno}, E.~{Bozzo}, N.~{Rea}, L.~{Pavan}, L.~{Burderi},
  M.~{Burgay}, S.~{Campana}, T.~{di Salvo}, M.~{Falanga}, M.D. {Filipovi{\'c}},
  P.C.C. {Freire}, J.W.T. {Hessels}, A.~{Possenti}, S.M. {Ransom}, A.~{Riggio},
  P.~{Romano}, J.M. {Sarkissian}, I.H. {Stairs}, L.~{Stella}, D.F. {Torres},
  M.H. {Wieringa}, G.F. {Wong}, Nature \textbf{501}(7468), 517 (2013).
\newblock \doi{10.1038/nature12470}

\bibitem{Altamiranoetal2010a}
D.~Altamirano, A.~Watts, M.~Linares, C.B. Markwardt, T.~Strohmayer, A.~Patruno,
  MNRAS \textbf{409}, 1136 (2010).
\newblock \doi{10.1111/j.1365-2966.2010.17369.x}

\bibitem{StrohmayerMarkwardt2010}
T.E. Strohmayer, C.B. Markwardt, The Astronomer's Telegram \textbf{2929}, 1
  (2010)

\bibitem{VerdhanChauhanetal2017}
J.V. Chauhan, J.S. Yadav, R.~Misra, P.C. Agrawal, H.M. Antia, M.~Pahari,
  N.~Sridhar, D.~Dedhia, T.~Katoch, P.~Madhwani, R.K. Manchanda, B.~Paul,
  P.~Shah, ApJ \textbf{841}, 41 (2017).
\newblock \doi{10.3847/1538-4357/aa6d7e}

\bibitem{Mahmoodifaretal2019}
S.~Mahmoodifar, T.E. Strohmayer, P.~Bult, D.~Altamirano, Z.~Arzoumanian,
  D.~Chakrabarty, K.C. Gendreau, S.~Guillot, J.~Homan, G.K. Jaisawal, L.~Keek,
  M.T. Wolff, ApJ \textbf{878}, 145 (2019).
\newblock \doi{10.3847/1538-4357/ab20c4}

\bibitem{Bultetal2019}
P.~Bult, G.K. Jaisawal, T.~Guver, T.E. Strohmayer, D.~Altamirano,
  Z.~Arzoumanian, D.R. Ballantyne, D.~Chakrabarty, J.~Chenevez, K.C. Gendreau,
  S.~Guillot, R.M. Ludlam, ApJ \textbf{885}, L1 (2019).
\newblock \doi{10.3847/2041-8213/ab4ae1}

\bibitem{Kaaretetal2007}
P.~Kaaret, Z.~Prieskorn, J.J.M. in't Zand, S.~Brandt, N.~Lund, S.~Mereghetti,
  D.~Gotz, E.~Kuulkers, J.A. Tomsick, ApJ \textbf{657}, L97 (2007).
\newblock \doi{10.1086/513270}

\bibitem{Bultetal2021}
P.~Bult, D.~Altamirano, Z.~Arzoumanian, A.V. Bilous, D.~Chakrabarty, K.C.
  Gendreau, T.~Guver, G.K. Jaisawal, E.~Kuulkers, C.~Malacaria, M.~Ng,
  A.~Sanna, T.E. Strohmayer, ApJ \textbf{907}, 79 (2021).
\newblock \doi{10.3847/1538-4357/abd54b}

\bibitem{Thompsonetal2005}
T.W.J. Thompson, R.E. Rothschild, J.A. Tomsick, H.L. Marshall, ApJ
  \textbf{634}, 1261 (2005).
\newblock \doi{10.1086/497104}

\bibitem{Strohmayeretal2008}
T.E. Strohmayer, C.B. Markwardt, E.~Kuulkers, ApJ \textbf{672}, L37 (2008).
\newblock \doi{10.1086/526546}

\bibitem{Foxetal2001}
D.W. Fox, W.H.G. Lewin, R.E. Rutledge, E.H. Morgan, R.~Guerriero, L.~Bildsten,
  M.~van~der Klis, J.~van Paradijs, C.B. Moore, T.~Dotani, K.~Asai, MNRAS
  \textbf{321}, 776 (2001).
\newblock \doi{10.1046/j.1365-8711.2001.04085.x}

\bibitem{Gallowayetal2001}
D.K. Galloway, D.~Chakrabarty, M.P. Muno, P.~Savov, ApJ \textbf{549}, L85
  (2001).
\newblock \doi{10.1086/319134}

\bibitem{Bhattacharyya2007}
S.~Bhattacharyya, MNRAS \textbf{377}, 198 (2007).
\newblock \doi{10.1111/j.1365-2966.2007.11587.x}

\bibitem{VillarrealStrohmayer2004}
A.R. Villarreal, T.E. Strohmayer, ApJ \textbf{614}, L121 (2004).
\newblock \doi{10.1086/425737}

\bibitem{Kaaretetal2003}
P.~Kaaret, J.J.M. in~'t Zand, J.~Heise, J.A. Tomsick, ApJ \textbf{598}, 481
  (2003).
\newblock \doi{10.1086/375053}

\bibitem{Altamiranoetal2008}
D.~Altamirano, P.~Casella, A.~Patruno, R.~Wijnands, M.~van~der Klis, ApJ
  \textbf{674}, L45 (2008).
\newblock \doi{10.1086/528983}

\bibitem{Munoetal2002}
M.P. Muno, D.~Chakrabarty, D.K. Galloway, D.~Psaltis, ApJ \textbf{580}, 1048
  (2002).
\newblock \doi{10.1086/343793}

\bibitem{Munoetal2000}
M.P. Muno, D.W. Fox, E.H. Morgan, L.~Bildsten, ApJ \textbf{542}, 1016 (2000).
\newblock \doi{10.1086/317031}

\bibitem{StrohmayerMarkwardt1999}
T.E. Strohmayer, C.B. Markwardt, ApJ \textbf{516}, L81 (1999).
\newblock \doi{10.1086/312009}

\bibitem{BhattacharyyaStrohmayer2005}
S.~Bhattacharyya, T.E. Strohmayer, ApJ \textbf{634}, L157 (2005).
\newblock \doi{10.1086/499100}

\bibitem{Munoetal2002a}
M.P. Muno, F.~Ozel, D.~Chakrabarty, ApJ \textbf{581}, 550 (2002).
\newblock \doi{10.1086/344152}

\bibitem{ChakrabortyBhattacharyya2014}
M.~Chakraborty, S.~Bhattacharyya, ApJ \textbf{792}, 4 (2014).
\newblock \doi{10.1088/0004-637X/792/1/4}

\bibitem{Strohmayeretal1997a}
T.E. Strohmayer, W.~Zhang, J.H. Swank, ApJ \textbf{487}, L77 (1997).
\newblock \doi{10.1086/310880}

\bibitem{Mahmoodifaretal2016}
S.~Mahmoodifar, T.E. Strohmayer, ApJ \textbf{818}, 93 (2016).
\newblock \doi{10.3847/0004-637X/818/1/93}

\bibitem{Zhangetal2013}
G.~Zhang, M.~M\'endez, T.~Belloni, J.~Homan, MNARS \textbf{436}, 2276 (2013).
\newblock \doi{10.1093/mnras/stt1728}

\bibitem{Zhangetal2016}
G.~Zhang, M.~M\'endez, M.~Zamfir, A.~Cumming, MNARS \textbf{455}, 2004 (2016).
\newblock \doi{10.1093/mnras/stv2482}

\bibitem{Ootesetal2017}
L.S. Ootes, A.L. Watts, D.K. Galloway, R.~Wijnands, ApJ \textbf{834}, 21
  (2017).
\newblock \doi{10.3847/1538-4357/834/1/21}

\bibitem{Munoetal2004}
M.P. Muno, D.K. Galloway, D.~Chakrabarty, ApJ \textbf{608}, 930 (2004).
\newblock \doi{10.1086/420812}

\bibitem{Munoetal2003}
M.P. Muno, F.~Ozel, D.~Chakrabarty, ApJ \textbf{595}, 1066 (2003).
\newblock \doi{10.1086/377447}

\bibitem{Chakrabortyetal2017}
M.~Chakraborty, Y.E. Bahar, E.~Gogus, ApJ \textbf{851}, 79 (2017).
\newblock \doi{10.3847/1538-4357/aa984e}

\bibitem{LeeMiller1998}
H.C. Lee, G.S. Miller, MNRAS \textbf{299}, 479 (1998).
\newblock \doi{10.1046/j.1365-8711.1998.01842.x}

\bibitem{Ford1999}
E.C. Ford, ApJ \textbf{519}, L73 (1999).
\newblock \doi{10.1086/312108}

\bibitem{Cuietal1998}
W.~Cui, E.H. Morgan, L.G. Titarchuk, ApJ \textbf{504}, L27 (1998).
\newblock \doi{10.1086/311569}

\bibitem{SazonovSunyaev2001}
S.Y. Sazonov, R.A. Sunyaev, A\&A \textbf{373}, 241 (2001).
\newblock \doi{10.1051/0004-6361:20010624}

\bibitem{Artigueetal2013}
R.~Artigue, D.~Barret, F.K. Lamb, K.H. Lo, M.C. Miller, MNRAS \textbf{433}, L64
  (2013).
\newblock \doi{10.1093/mnrasl/slt059}

\bibitem{Bhattacharyyaetal2018}
S.~Bhattacharyya, J.S. Yadav, N.~Sridhar, J.V. Chauhan, P.C. Agrawal, H.M.
  Antia, M.~Pahari, R.~Misra, T.~Katoch, R.K. Manchanda, B.~Paul, ApJ
  \textbf{860}, 2 (2018).
\newblock \doi{10.3847/1538-4357/aac495}

\bibitem{Wattsetal2005}
A.L. Watts, T.E. Strohmayer, C.B. Markwardt, ApJ \textbf{634}, 547 (2005).
\newblock \doi{10.1086/496953}

\bibitem{Wattsetal2008}
A.L. Watts, A.~Patruno, M.~van~der Klis, ApJ \textbf{688}, L37 (2008).
\newblock \doi{10.1086/594365}

\bibitem{BhattacharyyaStrohmayer2006}
S.~Bhattacharyya, T.E. Strohmayer, ApJ \textbf{642}, L161 (2006).
\newblock \doi{10.1086/504841}

\bibitem{WattsStrohmayer2006}
A.L. Watts, T.E. Strohmayer, MNRAS \textbf{373}, 769 (2006).
\newblock \doi{10.1111/j.1365-2966.2006.11072.x}

\bibitem{Chambersetal2018}
F.R.N. Chambers, A.L. Watts, Y.~Cavecchi, F.~Garcia, L.~Keek, MNRAS
  \textbf{477}, 4391 (2018).
\newblock \doi{10.1093/mnras/sty895}

\bibitem{Shara1982}
M.M. Shara, ApJ \textbf{261}, 649 (1982).
\newblock \doi{10.1086/160376}

\bibitem{CummingBildsten2000}
A.~Cumming, L.~Bildsten, ApJ \textbf{544}, 453 (1998).
\newblock \doi{10.1086/317191}

\bibitem{CooperNarayan2007}
R.L. Cooper, R.~Narayan, ApJ \textbf{657}, L29 (2007).
\newblock \doi{10.1086/513077}

\bibitem{FryxellWoosley1982b}
B.A. Fryxell, S.E. Woosley, ApJ \textbf{261}, 332 (1982).
\newblock \doi{10.1086/160344}

\bibitem{FryxellWoosley1982a}
B.A. Fryxell, S.E. Woosley, ApJ \textbf{258}, 733 (1982).
\newblock \doi{10.1086/160121}

\bibitem{Zingaleetal2001}
M.~Zingale, F.X. Timmes, B.~Fryxell, D.Q. Lamb, K.~Olson, A.C. Calder, L.J.
  Dursi, P.~Ricker, R.~Rosner, P.~MacNeice, H.M. Tufo, ApJS \textbf{133}, 195
  (2001).
\newblock \doi{10.1086/319182}

\bibitem{Cavecchietal2013}
Y.~Cavecchi, A.L. Watts, J.~Braithwaite, Y.~Levin, MNRAS \textbf{434}, 3526
  (2013).
\newblock \doi{10.1093/mnras/stt1273}

\bibitem{Cavecchietal2015}
Y.~Cavecchi, A.L. Watts, Y.~Levin, J.~Braithwaite, MNRAS \textbf{448}, 445
  (2015).
\newblock \doi{10.1093/mnras/stu2764}

\bibitem{Cavecchietal2016}
Y.~Cavecchi, Y.~Levin, A.L. Watts, J.~Braithwaite, MNRAS \textbf{459}, 1259
  (2016).
\newblock \doi{10.1093/mnras/stw728}

\bibitem{CavecchiSpitkovsky2019}
Y.~Cavecchi, S.~A, ApJ \textbf{882}, 142 (2019).
\newblock \doi{10.3847/1538-4357/ab3650}

\bibitem{Eidenetal2020}
K.~Eiden, M.~Zingale, A.~Harpole, D.~Willcox, Y.~Cavecchi, M.P. Katz, ApJ
  \textbf{894}, 6 (2020).
\newblock \doi{10.3847/1538-4357/ab80bc}

\bibitem{Bhattacharyyaetal2010}
S.~Bhattacharyya, M.C. Miller, D.K. Galloway, MNRAS \textbf{401}, 2 (2010).
\newblock \doi{10.1111/j.1365-2966.2009.15632.x}

\bibitem{vanStraatenetal2001}
S.~van Straaten, M.~van~der Klis, E.~Kuulkers, M.~M\'endez, ApJ \textbf{551},
  907 (2001).
\newblock \doi{10.1086/320234}

\bibitem{BhattacharyyaStrohmayer2007a}
S.~Bhattacharyya, T.E. Strohmayer, ApJ \textbf{656}, 414 (2007).
\newblock \doi{10.1086/510359}

\bibitem{BhattacharyyaStrohmayer2006a}
S.~Bhattacharyya, T.E. Strohmayer, ApJ \textbf{636}, L121 (2006).
\newblock \doi{10.1086/500199}

\bibitem{BhattacharyyaStrohmayer2006b}
S.~Bhattacharyya, T.E. Strohmayer, ApJ \textbf{641}, L53 (2006).
\newblock \doi{10.1086/503768}

\bibitem{BhattacharyyaStrohmayer2007}
S.~Bhattacharyya, T.E. Strohmayer, ApJ \textbf{666}, L85 (2007).
\newblock \doi{10.1086/521790}

\bibitem{MaurerWatts2008}
I.~Maurer, A.L. Watts, MNRAS \textbf{383}, 387 (2008).
\newblock \doi{10.1111/j.1365-2966.2007.12558.x}

\bibitem{Heyl2004}
J.S. Heyl, ApJ \textbf{600}, 939 (2004).
\newblock \doi{10.1086/379966}

\bibitem{Garciaetal2019}
F.~Garcia, F.R.N. Chambers, A.L. Watts, PhRvF \textbf{4}, 074802 (2019).
\newblock \doi{10.1103/PhysRevFluids.4.074802}

\bibitem{LeeStrohmayer2005}
U.~Lee, T.E. Strohmayer, MNRAS \textbf{361}, 659 (2005).
\newblock \doi{10.1111/j.1365-2966.2005.09198.x}

\bibitem{ChambersWatts2020}
F.R.N. Chambers, A.L. Watts, MNRAS \textbf{491}, 6032 (2020).
\newblock \doi{10.1093/mnras/stz3449}

\bibitem{Worpeletal2013}
H.~Worpel, D.K. Galloway, D.J. Price, ApJ \textbf{772}, 94 (2013).
\newblock \doi{10.1088/0004-637X/772/2/94}

\bibitem{GendreauArzoumanian2017}
K.~Gendreau, Z.~Arzoumanian, Nature Astronomy \textbf{1}, 895 (2017).
\newblock \doi{10.1038/s41550-017-0301-3}

\bibitem{Antiaetal2017}
H.M. Antia, J.S. Yadav, P.C. Agrawal, J.V. Chauhan, R.K. Manchanda, V.~Chitnis,
  B.~Paul, D.~Dedhia, P.~Shah, V.M. Gujar, T.~Katoch, V.N. Kurhade,
  P.~Madhwani, T.K. Manojkumar, V.A. Nikam, A.S. Pandya, J.V. Parmar, D.M.
  Pawar, M.~Pahari, R.~Misra, K.H. Navalgund, R.~Pandiyan, K.S. Sharma,
  K.~Subbarao, ApJS \textbf{231}, 10 (2017).
\newblock \doi{10.3847/1538-4365/aa7a0e}

\bibitem{Ghoshetal2013}
P.~Ghosh, L.~Angelini, M.~Baring, W.~Baumgartner, K.~Black, J.~Dotson,
  A.~Harding, J.~Hill, K.~Jahoda, P.~Kaaret, T.~Kallman, H.~Krawczynski,
  J.~Krolik, D.~Lai, C.~Markwardt, H.~Marshall, J.~Martoff, R.~Morris,
  T.~Okajima, R.~Petre, J.~Poutanen, S.~Reynolds, J.~Scargle, J.~Schnittman,
  P.~Serlemitsos, Y.~Soong, T.~Strohmayer, J.~Swank, Y.~Tawara, T.~Tamagawa,
  arXiv e-prints p. arXiv:1301.5514 (2013)

\bibitem{MillerLamb1998}
M.C. Miller, F.K. Lamb, ApJ \textbf{499}, L37 (1998).
\newblock \doi{10.1086/311335}

\bibitem{Nathetal2002}
N.R. Nath, T.E. Strohmayer, J.H. Swank, ApJ \textbf{564}, 353 (2002).
\newblock \doi{10.1086/324132}

\bibitem{Loetal2013}
K.H. Lo, M.C. Miller, S.~Bhattacharyya, F.K. Lamb, ApJ \textbf{776}, 19 (2013).
\newblock \doi{10.1088/0004-637X/776/1/19}

\end{thebibliography}

\end{document}